\documentclass[11pt]{scrartcl}
\usepackage[nodisplayskipstretch]{setspace}
\usepackage{ifxetex,ifluatex}
\usepackage{fixltx2e} 
\usepackage{etoolbox}
\usepackage{tcolorbox}
\tcbuselibrary{skins,breakable}
\IfFileExists{upquote.sty}{\usepackage{upquote}}{}
\IfFileExists{microtype.sty}{%
\usepackage{microtype}
\UseMicrotypeSet[protrusion]{basicmath} 
}{}
\usepackage{hyperref}
\hypersetup{unicode=true,
            pdftitle={eAppendix},
            pdfborder={0 0 0},
            breaklinks=false}
\usepackage{color}
\usepackage{fancyvrb}

\DefineVerbatimEnvironment{Highlighting}{Verbatim}{commandchars=\\\{\}}
\usepackage{framed}
\definecolor{shadecolor}{RGB}{248,248,248}
\newenvironment{Shaded}{\begin{snugshade}}{\end{snugshade}}

\newcommand{\CommentTok}[1]{\textcolor[rgb]{0.56,0.35,0.01}{\textit{#1}}}

\newcommand{\ControlFlowTok}[1]{\textcolor[rgb]{0.13,0.29,0.53}{\textbf{#1}}}
\newcommand{\DataTypeTok}[1]{\textcolor[rgb]{0.13,0.29,0.53}{#1}}
\newcommand{\DecValTok}[1]{\textcolor[rgb]{0.00,0.00,0.81}{#1}}

\newcommand{\FloatTok}[1]{\textcolor[rgb]{0.00,0.00,0.81}{#1}}

\newcommand{\KeywordTok}[1]{\textcolor[rgb]{0.13,0.29,0.53}{\textbf{#1}}}
\newcommand{\NormalTok}[1]{#1}
\newcommand{\OperatorTok}[1]{\textcolor[rgb]{0.81,0.36,0.00}{\textbf{#1}}}
\newcommand{\OtherTok}[1]{\textcolor[rgb]{0.56,0.35,0.01}{#1}}

\newcommand{\SpecialStringTok}[1]{\textcolor[rgb]{0.31,0.60,0.02}{#1}}
\newcommand{\StringTok}[1]{\textcolor[rgb]{0.31,0.60,0.02}{#1}}

\usepackage{fancyvrb}
\usepackage{fvextra}

\usepackage{graphicx,grffile}
\makeatletter
\def\maxwidth{\ifdim\Gin@nat@width>\linewidth\linewidth\else\Gin@nat@width\fi}
\def\maxheight{\ifdim\Gin@nat@height>\textheight\textheight\else\Gin@nat@height\fi}
\makeatother
\setkeys{Gin}{width=\maxwidth,height=\maxheight,keepaspectratio}

\ifx\paragraph\undefined\else
\let\oldparagraph\paragraph
\renewcommand{\paragraph}[1]{\oldparagraph{#1}\mbox{}}
\fi
\ifx\subparagraph\undefined\else
\let\oldsubparagraph\subparagraph
\renewcommand{\subparagraph}[1]{\oldsubparagraph{#1}\mbox{}}
\fi

\let\rmarkdownfootnote\footnote%
\def\footnote{\protect\rmarkdownfootnote}

\usepackage [autostyle, english = american]{csquotes}
\MakeOuterQuote{"}

\usepackage{xcolor}
\usepackage{textcomp}
\usepackage{booktabs}
\usepackage[margin=1in, footskip=.5in]{geometry}
\usepackage{float}
\usepackage{lipsum}
\usepackage{wrapfig}
\usepackage{enumitem}
\usepackage{amsmath}
\usepackage[colorinlistoftodos]{todonotes}
\usepackage{pgf,tikz}
\usepackage{tkz-tab}
\usepackage[linewidth = .5pt]{mdframed}
\usepackage{multirow}
\usepackage{nicefrac}
\usepackage[nocompress]{cite}
\usepackage{amssymb}
\usepackage{amsbsy}
\usepackage{bm}
\usepackage{array}
\usepackage{anyfontsize}
\usepackage{bold-extra}
\usepackage{bibentry}
\usepackage{longtable}
\usepackage{pdflscape}
\usepackage[labelfont=bf]{caption}
\usepackage{ragged2e}
\usetikzlibrary{positioning, calc, shapes.geometric, shapes.multipart,
  shapes, arrows.meta, arrows,
  decorations.markings, external, trees}
\usepackage{tkz-tab}
\usepackage{relsize}

\allowdisplaybreaks
\usepackage[all]{nowidow}

\setkomafont{title}{\normalfont\upshape\bfseries}
\setkomafont{section}{\normalfont\large\upshape\bfseries}
\addtokomafont{section}{\MakeUppercase}
\setkomafont{subsection}{\normalfont\upshape\bfseries}
\setkomafont{subsubsection}{\normalfont\upshape\itshape}
\setlist[enumerate]{leftmargin=*}
\usetikzlibrary{positioning}
\usetikzlibrary{shapes}
\usetikzlibrary{arrows.meta}
\makeatletter
\renewcommand*\env@matrix[1][\arraystretch]{%
  \edef\arraystretch{#1}%
  \hskip -\arraycolsep
  \let\@ifnextchar\new@ifnextchar
  \array{*\c@MaxMatrixCols c}}
\makeatother

\usepackage{array}
\usepackage{amsmath}
\usepackage{ifxetex,ifluatex}
\ifxetex
  \usepackage{fontspec,xltxtra,xunicode}
  \defaultfontfeatures{Mapping=tex-text,Scale=MatchLowercase}
\else
  \ifluatex
    \usepackage{fontspec}
    \defaultfontfeatures{Mapping=tex-text,Scale=MatchLowercase}
  \else
    \usepackage[utf8]{inputenc}
  \fi
\fi

\definecolor{keyword}{HTML}{1b6ca2}
\definecolor{argument}{HTML}{2d9027}
\definecolor{flow}{HTML}{ff7f00}
\definecolor{number}{HTML}{e31a1c}
\definecolor{variable}{HTML}{b15928}
\definecolor{function}{HTML}{1f78b4}
\definecolor{text}{HTML}{4b4b4b}
\definecolor{string}{HTML}{6a3d9a}
\definecolor{comment}{HTML}{a2a2a2}
\definecolor{operator}{HTML}{ff7f00}
\definecolor{logical}{HTML}{e31a1c}
\definecolor{warning}{HTML}{e31a1c}
\definecolor{alert}{HTML}{e31a1c}
\definecolor{error}{HTML}{e31a1c}
\definecolor{shadecolor}{HTML}{f8f8f8}
\definecolor{sidecolor}{HTML}{717171}

\DefineShortVerb[commandchars=\\\{\}]{\|}
\DefineVerbatimEnvironment{Highlighting}{Verbatim}{breaklines,commandchars=\\\{\}, formatcom={\setstretch{1.1} \color{text}}}

\renewenvironment{Shaded}{\begin{tcolorbox}[
    breakable,
    colback=shadecolor,
    colframe=shadecolor,
    coltext=text,
    boxsep=-3pt,
    bottom=-8pt,
    oversize,
    sharp corners=all,
    before skip=12pt,
    after skip=12pt,
    height fixed for=first and middle
  ]%
  }{\end{tcolorbox}}
\fvset{breaksymbolleft={}, breakautoindent=true, breakindent=4ex}

\renewenvironment{verbatim}
 {\VerbatimEnvironment
  \begin{tcolorbox}[
  enhanced,
    breakable,
    colback=shadecolor,
    colframe=shadecolor,
    coltext=text,
    boxsep=-3pt,
    oversize,
    sharp corners=all,
    borderline west={2pt}{0pt}{sidecolor},
    after skip = 12pt,
    height fixed for=first and middle
  ]%
    \begin{Verbatim}[breaklines,formatcom={\setstretch{1.1}}]}
    {\end{Verbatim}\end{tcolorbox}}
\renewcommand{\KeywordTok}[1]{\textcolor{keyword}{#1}}
\renewcommand{\DataTypeTok}[1]{\textcolor{argument}{#1}}
\renewcommand{\DecValTok}[1]{\textcolor{number}{#1}}

\renewcommand{\FloatTok}[1]{\textcolor{number}{#1}}

\renewcommand{\StringTok}[1]{\textcolor{string}{#1}}

\renewcommand{\SpecialStringTok}[1]{\textcolor{string}}

\renewcommand{\CommentTok}[1]{\textcolor{comment}{\textit{#1}}}

\renewcommand{\OtherTok}[1]{\textcolor{logical}{#1}}

\renewcommand{\ControlFlowTok}[1]{\textcolor{flow}{\textbf{#1}}}
\renewcommand{\OperatorTok}[1]{\textcolor{operator}{\textbf{#1}}}

\renewcommand{\NormalTok}[1]{\textcolor{text}{#1}}

\renewcommand\paragraph{\newpage}

\newcommand{\RR}[1]{\text{RR}_{#1}}
\newcommand{\OR}[1]{\text{OR}_{#1}}
\newcommand{\Prc}[2]{\Pr\left({#1}\mid{#2}\right)}
\newcolumntype{P}[1]{>{\RaggedRight\hspace{0pt}}p{#1}}

\newcommand*{\TitleFont}{%
      \fontsize{18}{12}%
      \selectfont
      \centering}
\newcommand*{\AuthFont}{%
      \fontsize{14}{12}%
      \selectfont
      \centering}
\newcommand*{\InstFont}{%
      \fontsize{12}{12}%
      \selectfont
      \centering}

\usepackage{stringenc}

\newcommand*{\ditto}{\texttt{\char`\"}}

\usepackage{indentfirst}

\begin{document}

\begin{center}
  \vspace{-4ex}
  \TitleFont{
  Multiple-bias sensitivity analysis using bounds
  }

  \AuthFont{
  Louisa H. Smith\textsuperscript{a}, Maya B. Mathur\textsuperscript{b}, Tyler J. VanderWeele\textsuperscript{a}
  }
  \vspace{-.5ex}

 \InstFont{
\textsuperscript{a}Harvard T.H. Chan School of Public Health, \textsuperscript{b}Stanford University
}

\end{center}

\begin{abstract}
Unmeasured confounding, selection bias, and measurement error are well-known sources of bias in epidemiologic research. Methods for assessing these biases have their own limitations. Many quantitative sensitivity analysis approaches consider each type of bias individually, while more complex approaches are harder to implement or require numerous assumptions. By failing to consider multiple biases at once, researchers can underestimate -- or overestimate -- their joint impact. We show that it is possible to bound the \emph{total} composite bias due to these three sources, and to use that bound to assess the sensitivity of a risk ratio to any combination of these biases. We derive bounds for the total composite bias under a variety of scenarios, providing researchers with tools to assess their total potential impact. We apply this technique to a study where unmeasured confounding and selection bias are both concerns, and to another study in which possible differential exposure misclassification and unmeasured confounding are concerns. We also show that a \ditto multi-bias E-value\ditto  can describe the minimal strength of joint bias-parameter association necessary for an observed risk ratio to be compatible with a null causal effect (or with other pre-specified effect sizes). This may provide intuition about the relative impacts of each type of bias. The approach we describe is easy to implement with minimal assumptions, and we provide R functions to do so.
\end{abstract}

\thispagestyle{empty}
\clearpage
\pagenumbering{arabic}

\newpage

\hypertarget{introduction}{%
\section*{Introduction}\label{introduction}}

Assessing evidence for causation is fundamental in order to plan and target interventions and improve public health. However, many causal claims in epidemiologic studies are met with suspicion by both researchers and the general public, due to the fact that such studies are well known to be subject to various biases. While faults in these studies can sometimes be addressed directly -- e.g., through better sampling schemes, blinded outcome ascertainment, more extensive covariate measurements, etc. -- other times unmeasured confounding, selection bias, and measurement error are unavoidable. In such situations, our next best option is to assess the extent to which a given study's conclusions might be sensitive or robust to these biases, and whether they threaten its conclusions. Often, however, this is limited to a few sentences in a discussion section qualitatively assessing the possibility of bias, sometimes appealing without quantitative justification to heuristics that may or may not hold true in a particular study.\textsuperscript{1,2}

The weak uptake of quantitative bias analysis in epidemiology belies its long history. Over a half-century ago, Cornfield and then Bross argued that the extent of possible bias was quantifiable based on observed data and possibly hypothetical quantities.\textsuperscript{3--6} Attempts to generalize these results as well as consider other biases, sometimes simultaneously with confounding, followed.\textsuperscript{6--11} More recently, probabilistic bias analysis methods have been developed, allowing researchers to propose distributions for various bias parameters across multiple biases, and to explore how various combinations of those parameters would affect their results.\textsuperscript{12--17} Despite the availability of these methods in textbook and software form,\textsuperscript{15,18,19} the actual uptake of such quantitative bias analysis in empirical research has been limited,\textsuperscript{20} possibly because of the (at least perceived) computational complexity\textsuperscript{21} or difficulty in proposing plausible distributions.

Even more recently, simpler approaches to sensitivity analysis have hearkened back to the early days of bias assessment, with the development of bounds for various biases that require minimal assumptions and at most basic algebra.\textsuperscript{22--24} The E-value, a consequence of the bound for unmeasured confounding,\textsuperscript{25} has proved to be a popular way to conceive of that bias,\textsuperscript{26} and equivalent quantities are available for other biases.\textsuperscript{23,24} However, a one-at-a-time approach is not sufficient for many studies subject to multiple sources of bias. In this article we extend the simple sensitivity analysis framework to multiple biases, describing a bound for the total composite bias from unmeasured confounding, selection, and differential exposure or outcome misclassification.

\hypertarget{the-problem-of-multiple-biases}{%
\section*{The problem of multiple biases}\label{the-problem-of-multiple-biases}}

We will describe a scenario in which all three types of bias are present, preventing the causal interpretation of an observed risk ratio as a causal risk ratio. Consider a binary exposure \(A\), a binary outcome \(Y\), and measured covariates \(C\). Let \(S\) be an indicator of the subset of the population for which data has been collected, and let \(A^*\) and \(Y^*\) denote misclassified versions of the exposure and outcome, respectively. We use potential outcome notation to describe causal quantities: \(Y_a\) is the outcome that would occur were exposure \(A\) set to value \(a\). We assume consistency, meaning that \(Y_a = Y\) for observations for whom we observe \(A = a\), and positivity, meaning that \(0 < \Prc{A = 1}{\cdot} < 1\) within every stratum of the population.

We denote (conditional) independence between random variables with the symbol \(\amalg\), such that \(Y_a \amalg A \mid C\) implies conditional exchangeability; i.e., potential outcomes are independent of exposure status conditional on \(C\). However, when \(C\) does not capture all of the exposure-outcome confounding, it is not true that \(Y_a \amalg A \mid C\). We assume in that case that additionally adjusting for some unmeasured factor(s) \(U_c\) would be sufficient to address confounding, so that \(Y_a \amalg A \mid C, U_c\). \(U_c\) may be a single random variable or a vector of variables, which may be continuous or take on any number of discrete values, or some combination. Similarly, we allow for selection bias, which we define as a lack of the conditional independence \(Y_a \amalg A \mid C, U_c, S = 1\). We likewise assume that the measurement of some variable(s) \(U_s\), responsible for selection, would fully account for this bias, though the necessary conditions for it to do so will depend on whether we intend to make inferences about effects in the total population, or just the selected population. Finally, we allow for the possibility that the misclassification is differential, by which we mean that the sensitivity and specificity of the exposure measurement may differ depending on the value of the outcome, or that the sensitivity and specificity of the outcome measurement may depend on the exposure. In our notation, this means that it is not necessarily true that \(A^* \amalg Y \mid A, C\) or that \(Y^* \amalg A \mid Y, C\).

\hypertarget{motivating-examples}{%
\subsection*{Motivating examples}\label{motivating-examples}}

There is great interest in how exposures during pregnancy may affect offspring health. However important such questions are, they are difficult to answer with epidemiologic research. Ethics may limit inclusion of pregnant people in randomized trials, and many exposures of interest are not ethical or feasible to randomize to anyone. Case-control studies can efficiently capture rare childhood outcomes, but recalling pregnancy exposures several years later can result in measurement error.\textsuperscript{27} Prospective cohort studies can avoid this recall bias, but are often subject to loss to follow-up when the duration between exposure and outcome assessment is long.\textsuperscript{28} Observational studies of all types are threatened by unmeasured confounding, and inter-generational confounders are particularly difficult to assess.\textsuperscript{29} Importantly, studies like these are not affected by only one or another of these biases, but may suffer from multiple threats to validity.

To demonstrate our sensitivity analysis approach, we will consider two questions about exposures during pregnancy and outcomes in children: whether HIV infection in utero causes wasting (low weight-for-length), and whether vitamin consumption during pregnancy protects against childhood leukemia.

Omoni et al.~investigated the former hypothesis in participants of a vitamin A supplementation trial in Zimbabwe and found that, compared to children who were unexposed to HIV, those who had been infected with HIV in utero were significantly more likely to be below a weight-for-length Z-score of -2 as toddlers.\textsuperscript{30} The odds ratio comparing the two groups was 6.75 (95\% CI, 2.79, 16.31) at 2 years. Although randomized trial data were used for the analysis, this was an observational study with respect to HIV infection, since infection is not randomized. The authors did not, however, adjust for any confounders. Furthermore, since enrollment occurred at delivery, after possible HIV exposure and transmission, the choice of whether to participate could have been affected by HIV status as well as other factors. We will consider the role that confounding and selection bias may play in this study.

As a second example, Ross et al.~analyzed the relationship between vitamins and leukemia in a case-control study and found a decreased risk of acute lymphoblastic leukemia among children whose mothers consumed vitamin supplements during pregnancy.\textsuperscript{31} Their reported odds ratio, which, with a rare outcome, approximates a risk ratio, of 0.51 (95\% CI 0.30, 0.89) was conditional on maternal age, race, and a binary indicator of education. However, there may be other confounders that were not adjusted for, such as other indicators of a privileged or healthy lifestyle that are both associated with vitamin use and protective against leukemia. We also may be concerned about recall bias -- that mothers of children with a cancer diagnosis might be more likely to report \emph{not} taking a vitamin even if they did so -- so we consider how exposure misclassification and unmeasured confounding can be assessed simultaneously.

\hypertarget{the-multiple-bias-bound}{%
\section*{The multiple-bias bound}\label{the-multiple-bias-bound}}

Two overarching types of bias analysis have been described: one that explores how biases of a given magnitude affect an estimate, which Phillips labeled \enquote{bias-level sensitivity analysis}, and another that reduces the analysis to a summary of how much bias would be necessary for an observation to be compatible with a truly null effect (or some other specified non-null effect), which he called \enquote{target-adjusted sensitivity analysis.}\textsuperscript{32} Our approach allows for both. First we consider the multiple-bias bound, which allows researchers or consumers of research to explore the maximum factor by which unmeasured confounding, selection, and misclassification could bias a risk ratio.

We begin with outcome misclassification, and then extend our results to exposure misclassification. We assume that the investigators have estimated \(\text{RR}^{\text{obs}}_{AY^*}= \frac{\Prc{Y^* = 1}{A = 1, S = 1, c}}{\Prc{Y^* = 1}{A = 0, S = 1, c}}\) but wish to make inference about \(\text{RR}^{\text{true}}_{AY}= \frac{\Prc{Y_1 = 1}{c}}{\Prc{Y_0 = 1}{c}}\). We will assess bias on the relative scale, so that we define the bias as \(\text{RR}^{\text{obs}}_{AY^*}/ \text{RR}^{\text{true}}_{AY}\).

Using bounds that have been previously described for misclassification, selection bias, and unmeasured confounding considered individually,\textsuperscript{22--25} we can bound \(\text{RR}^{\text{obs}}_{AY^*}\) by factoring it into \(\text{RR}^{\text{true}}_{AY}\) and components for each of the biases. The parameters that will be used to bound the biases are as follows:
\[
\RR{AY^* \mid y, S = 1} = \max_y{\frac{\Prc{Y^* = 1}{Y = y, A = 1, S = 1, c}}{\Prc{Y^* = 1}{Y = y, A = 0, S = 1, c}}} \]
\[
\RR{U_sY\mid A = a}  = \frac{\max_u \Prc{Y = 1}{A = a, c, U_s = u}}{\min_u \Prc{Y = 1}{A = a, c, U_s = u}}\;\; \text{for } a = 0, 1
\]
\[
\RR{SU_s\mid A = a}  = \max_u\frac{\Prc{U_s = u}{A = a, S = a, c}}{\Prc{U_s = u}{A = a, S = 1 - a, c}} \;\; \text{for } a = 0, 1 
\]
\[
\RR{U_cY }  = \max_a\frac{\max_u \Prc{Y = 1}{A = a, c, U_c = u}}{\min_u \Prc{Y = 1}{A = a, c, U_c = u}} 
\]
\[
\RR{AU_c}  = \max_u \frac{\Prc{U_c = u}{A = 1, c}}{\Prc{U_c = u}{A = 0, c}} \;\; .
\]

These bias parameters have been described elsewhere, though separately.\textsuperscript{22--24} Briefly, the bias parameter defining the misclassification portion of the bound describes the maximum of the false positive probability ratio or sensitivity ratio \emph{within} the selected population. The selection bias parameters describe the maximum factors by which the outcome risk differs by values of \(U_s\), within strata of \(A\), and the maximum factors by which some level of \(U_s\) differs between the selected and non-selected groups, within strata of \(A\). Finally, the unmeasured confounding parameters describe the maximum factor by which \(U_c\) increases the outcome risk, conditional on \(A\), and the maximum factor by which exposure is associated with some value of \(U_c\). Each of the sensitivity parameters is conditional on the covariates adjusted for in the analysis, and so describes the extent of bias above and beyond those factors.

To simplify notation, define the function \(g(a, b) = \frac{a \times b}{a + b - 1}\). Then we have the following bound for the total composite bias.

\textbf{Result 1}:

If \(Y_a \amalg A \mid C, U_c\) and \(Y \amalg S \mid A, C, U_s\), then:
\[
\text{RR}^{\text{obs}}_{AY^*}/ \text{RR}^{\text{true}}_{AY}\leq \text{BF}_{m}\times \text{BF}_{s}\times \text{BF}_{c}
\]
where \(\text{BF}_{m}= \RR{AY^* \mid y, S = 1}\), \(\text{BF}_{s}= g\left(\RR{U_sY\mid A = 1}, \RR{SU_s\mid A = 1}\right) \times g\left(\RR{U_sY\mid A = 0}, \RR{SU_s\mid A = 0}\right)\), and \(\text{BF}_{c}= g\left(\RR{AU_c}, \RR{U_cY} \right)\). The derivation of this and the results that follow are given in the eAppendix.

Result 1 can be used to quantify the maximum amount of bias that could be produced by parameters of a given value. Values for the sensitivity parameters may be taken from validation studies, previous literature, or expert knowledge, or proposed as hypotheticals. Because the sensitivity parameters are maxima they are always greater than or equal to 1 and the composite bound will thus be greater than or equal to 1. For an apparently causative observed exposure-outcome risk ratio (\textgreater1) one could divide the estimate and its confidence interval by the bound to obtain the maximum that the specified biases could shift the estimate and its confidence interval. For a preventive observed exposure-outcome risk ratio (\textless1) one could multiply the estimate and its confidence interval by the bound to obtain the maximum that the specified biases could shift the estimate and its confidence interval. Although the bound allows for terms for all three biases, if any of them is judged not to threaten a given study, that factor can be omitted. Furthermore, the selection bias term can be simplified under certain assumptions;\textsuperscript{23} we illustrate in the first example below.

Note that the factorization of the bound implies an ordering of the biases: the misclassification parameters are defined within the stratum \(S = 1\). Intuitively, this corresponds to a study in which outcome measurement is done after people have been selected into the study, and so requires considering the strength of differential misclassification only within that group. In general, we can think of biases as layers that we must peel off sequentially, and the order in which we do so is the reverse of the order in which they occurred.\textsuperscript{33,34} Confounding is generally thought of as a property of nature within the population of interest, so occurs first (though if parameters describing the strength of confounding are derived based on misclassified exposure or outcome, that may not be the case\textsuperscript{33}), but the order in which selection and misclassification occur may depend on the study design. We could alternatively derive a bound that depends on a parameter describing the extent to which the outcome is misclassified in the total population, and on others describing how selection is associated with the misclassified outcome. These parameters may be more intuitive in a study with case-control sampling. We define the alternative parameters and derive that bound in the eAppendix.

\hypertarget{example}{%
\subsubsection{Example}\label{example}}

We can use the multiple-bias bound to assess possible bias in the study by Omoni and colleagues regarding the effect of HIV status on wasting.\textsuperscript{30} Wasting is defined by weight-for-length Z-score of -2 or below and is a rare outcome, so we can interpret the reported OR of 6.75 (95\% CI, 2.79, 16.31) as an approximate risk ratio. Since we have no reason to believe that misclassification of wasting was differential by exposure status (i.e., child or mother HIV status) status, we will focus on unmeasured confounding and selection bias in this example.

The choice of whether to participate in the trial, and therefore in the analysis in question, may have been influenced by prior maternal HIV status. For example, people with HIV infection may be hesitant to enroll due to stigma regarding infection, or fear of confirming their status. Other factors may affect enrollment as well: parents with food insecurity may be more likely to enroll in a vitamin-supplementation trial than those without, if they think it will improve their children's nutrition. Selection bias could result, therefore, if food insecurity also increases the risk of wasting. Similarly, there are factors that are associated with HIV status that may also affect wasting; if these are not on the causal pathway, we may be worried about unmeasured confounding. The authors did not adjust for parity or marital status, though they report that primiparous women were less likely to have HIV, as were married women. We may be concerned that children in single-parent households and those with more siblings are at higher risk of wasting. A directed acyclic graph depicting these relationships is shown in Figure 1A.

Suppose that the most vulnerable in the population were more likely to participate in the trial, and thus that wasting is more likely in children of participants than of non-participants, both among those with HIV as well as those without. The assumption that the outcome is more likely in the selected population of both exposure groups allows us to simplify the bound, so that the bounding factor only relies on two terms. Suppose now that children of the most food-insecure mothers are 3 times as likely to have extremely low weight-for-length scores than the least likely group, so that \(\RR{U_sY \mid A = 1} = 3\), and that the mothers with HIV infection in the study compared to those not in the study are twice as likely to be food insecure, so that \(\RR{SU_s \mid A = 1} = 2\).

Although the odds ratios from this study were not adjusted for parity and marital status, the authors reported proportions of these characteristics stratified by exposure,\textsuperscript{30} which can aid in coming up with a reasonable value for \(\RR{AU_c}\). For example, suppose we estimate that 3\% of the women whose infants are infected with HIV are multiparous and unmarried, but that this is true of 7\% of the women without HIV. If this is the family situation with the largest disparity between exposure groups, then we can specify \(\RR{AU_c} = 2.3\). Now suppose that children in these most precarious families have 2.5 times the risk of wasting than those in the least precarious, so that \(\RR{U_cY} = 2.5\).

Then we can calculate the bound as \(\frac{3 \times 2}{3 + 2 - 1} \times \frac{2.3 \times 2.5}{2.3 + 2.5 - 1} = 2.27\). If those are the only sources of selection bias and unmeasured confounding, and there is no measurement error, then it is clear that this amount of bias cannot fully explain the approximate observed \(\text{RR}^{\text{obs}}_{AY}\) of 6.75. Of course, this observed value is subject to statistical uncertainty, so we can also consider the lower limit of the confidence interval, 2.79. If the proposed parameter values hold, then even in the worst case scenario, \(\text{RR}^{\text{true}}_{AY}\) is still consistent with \(2.79 / 2.27 = 1.23\), an increase of about 23\% in risk of wasting at 2 years of age due to HIV infection.

\hypertarget{exposure-misclassification}{%
\subsection*{Exposure misclassification}\label{exposure-misclassification}}

When differential exposure misclassification is a concern, we can derive a similar bound under similar assumptions. However, unlike the bound for outcome misclassification, the bound for exposure misclassification that is employed applies to the odds ratio, not the risk ratio, and the sensitivity parameters are also not themselves risk ratios. We therefore cannot factor the observed risk ratio as in the previous section. However, for a sufficiently rare outcome, odds ratios approximate risk ratios, which allows for some progress.

In this section, \(\text{RR}^{\text{obs}}_{A^*Y}= \frac{\Prc{Y = 1}{A^* = 1, S = 1, c}}{\Prc{Y = 1}{A^* = 0, S = 1, c}}\) refers to the observed (approximate) risk ratio under exposure misclassification, when the outcome is rare in the selected population. Denote with \(\OR{A^*Y \mid y, S = 1}\) the largest out of the false-positive odds ratio \(\left\{f'_1/f'_0\right\} / \left\{(1 - f'_1)/(1 - f'_0)\right\}\), the sensitivity odds ratio \(\left\{s'_1/s'_0\right\} / \left\{(1 - s'_1)/(1 - s'_0)\right\}\), the correct classification ratio \(\{s'_1/s'_0\} / \{(1 - f'_1)/(1 - f'_0)\}\), and incorrect classification ratio \(\left\{f'_1/f'_0\right\} / \left\{(1 - s'_1)/(1 - s'_0)\right\}\), where \(f'_y = \Pr(A^* = 1\mid Y = y, A = 0, S = 1, c)\) and \(s'_y = \Prc{A^* = 1}{Y = y, A = 1, S = 1, c}\). Then the following bound holds approximately.

\textbf{Result 2}:

If \(\Prc{Y = 0}{ A^* = a, S = 1, c} \approx 1\) and \(\Prc{Y = 0}{ A = a, S = 1, c} \approx 1\), then:
\[
\text{RR}^{\text{obs}}_{A^*Y}/ \text{RR}^{\text{true}}_{AY}\leq \text{BF}_{m}\times \text{BF}_{s}\times \text{BF}_{c}
\]
where \(\text{BF}_{m}= \OR{A^*Y \mid y, S = 1}\) and \(\text{BF}_{s}\) and \(\text{BF}_{c}\) are as previously defined.

\hypertarget{example-1}{%
\subsubsection{Example}\label{example-1}}

We can jointly assess the magnitude of bias due to differential recall of vitamin use and unmeasured confounding in the study of leukemia risk by Ross and colleagues, in which \(\text{RR}^{\text{obs}}_{A^*Y}= 0.51\) (95\% CI 0.30, 0.89), by proposing realistic values for the bias parameters. A probabilistic bias analysis for misclassification was previously done in relation to this study, in which Jurek et al.~conducted a literature search for validation studies of multivitamin use during the periconceptional period.\textsuperscript{35} They found no pertinent articles, and instead used expert knowledge and bounds from the data (e.g., by assuming correct classification is better than chance) to propose distributions for false negative and false positive probabilities for the cases and controls, which we can use to inform our choice of parameters. Because we think the case-control differential in false negatives is stronger than that for false positives, we might choose that \(\Prc{A^* = 0}{Y = 1, A = 1} = 0.15\) and \(\Prc{A^* = 0}{Y = 1, A = 1} = 0.1\), and \(\text{BF}_{m}' = 1.59\).

Jurek et al.'s probabilistic bias analysis used the raw 2-by-2 table from the original article, so did not take into account even the few measured confounders.\textsuperscript{35} However, even those measured confounders would likely not be sufficient to control for confounding by healthy lifestyle, as there is evidence that other healthy behaviors are associated with leukemia. For example, a recent meta-analysis found that not breastfeeding compared to breastfeeding for at least 6 months was associated with an increase in acute lymphoblastic leukemia risk by a factor of 1.22.\textsuperscript{36} Using breastfeeding as a proxy for healthy lifestyle, for the unmeasured confounding parameters, we will take \(\RR{U_cY} = 1.22\) and \(\RR{AU_c} = 2\), suggesting that children who weren't breastfed are 1.22 times as likely to get leukemia, and that mothers who take multivitamins are twice as likely to breastfeed than those who do not. A directed acyclic graph depicting this example is shown in Figure 1B.

Using these values, we find that \(1.59 \times \frac{1.22 \times 2}{1.22 + 2 - 1} = 1.75\), indicating that the observed risk ratio of 0.51 may be biased by a factor of 1.75 if the differential misclassification and unmeasured confounding were of the strengths we proposed. Since we are dealing with a possibly protective factor, we multiply the observed estimate of 0.51 and its confidence interval (95\% CI 0.30, 0.89) by the bound, resulting in a corrected estimate and confidence interval of 0.89 (95\% CI 0.52, 1.56). Unlike the Jurek et al.~sensitivity analysis,\textsuperscript{35} which found that results were largely unchanged by exposure misclassification, we have focused specifically on a situation in which misclassification is differential by outcome, and have additionally taken both measured and unmeasured confounding into account. Doing so indicates that the results may be sensitive to misclassification and unmeasured confounding, as can be seen if the chosen parameter values are thought to be reasonable.

\hypertarget{inference-in-the-selected-population}{%
\subsection*{Inference in the selected population}\label{inference-in-the-selected-population}}

Results 1 and 2 are derived with respect to the true causal effect in the total population, despite possible selection bias. In other situations, we may only be interested in the existence and magnitude of a causal effect in the selected population. In this case, our estimand of interest is \(\text{RR}^{\text{true}}_{AY\mid S = 1}= \frac{\Prc{Y_1}{S = 1, c}}{\Prc{Y_0}{S = 1, c}}\). If only selection bias is present, one can derive a bound under the assumption that \(Y_a \amalg A \mid S = 1, c, U_s\).\textsuperscript{23} In the present context, we additionally accommodate unmeasured confounding by \(U_c\) such that it is only the case that \(Y_a \amalg A \mid S = 1, c, U_s, U_c\). Therefore, we must consider the vector of factors causing selection bias and unmeasured confounding \(U_{sc} = \left(U_s, U_c\right)\). Define the sensitivity parameters \(\RR{U_{sc}Y} = \max_a\frac{\max_u \Prc{Y = 1}{A = a, c, U_{sc} = u}}{\min_u \Prc{Y = 1}{A = a, c, U_{sc} = u}}\) and \(\RR{AU_{sc}} = \max_u \frac{\Prc{U_{sc} = u}{A = 1, c}}{\Prc{U_{sc} = u}{A = 0, c}}\).

Then under outcome misclassification, we have the following bound.

\textbf{Result 3}:

If \(Y_a \amalg A \mid S = 1, c, U_c ,U_s\), then:
\[
\text{RR}^{\text{obs}}_{AY^*}/ \text{RR}^{\text{true}}_{AY\mid S = 1}\leq \text{BF}_{m}\times \text{BF}_{sc}
\]
where \(\text{BF}_{m}\) is defined as in Result 1, and \(\text{BF}_{sc}= g\left(\RR{U_{sc}Y},\RR{AU_{sc}}\right)\). These latter parameters now refer to the maximum risk ratio for the outcome among the selected comparing any two levels of any of \(U_s\) and \(U_c\), and the maximum ratio for any joint level of \(U_s\) and \(U_c\) comparing exposed to unexposed, among the selected. This bound holds under exposure misclassification with a rare outcome in the selected population as well, with \(\text{BF}_{m}' = \OR{A^*Y \mid y, S = 1}\).

A summary of all results, including those with a different ordering of selection and misclassification, and those targeting the selected population, is given in Table 1. The multi-bias E-value, described in the next section, is given in the final column of Table 1.

\hypertarget{the-multi-bias-e-value}{%
\section*{The multi-bias E-value}\label{the-multi-bias-e-value}}

The bounds in Results 1, 2, and 3 allow researchers and consumers of research to choose values for bias parameters and investigate their possible effects on an observed risk ratio. Target-adjusted sensitivity analysis, on the other hand, quantifies the strength of bias necessary to shift an observation to another value, often the null value, though others can be used.\textsuperscript{37} The E-value for unmeasured confounding is an example of this approach.\textsuperscript{25} We can calculate an equivalent value for a combination of biases using the bounds in this article. The E-value for unmeasured confounding refers to a value that can be shown to be sufficient to explain away an observed estimate and that jointly minimizes the maximum of the two sensitivity parameters for unmeasured confounding.\textsuperscript{25} Similarly, the multi-bias E-value describes the minimum value that all of the sensitivity parameters for each of the biases would have to take on for a given observed risk ratio to be compatible with a truly null risk ratio. Since the overall bias is monotone increasing in the individual bias parameters, it follows that if any one of the bias parameters is less than the multi-bias E-value, then at least one other parameter would have to be greater than the multi-bias E-value in order to completely explain a result.

Recall that under non-differential misclassification of the exposure, the \(\text{BF}_{m}\) factor in the bound is not a risk ratio. If the misclassified exposure is rare, then that parameter can be interpreted as an approximate risk ratio; otherwise, an approximate square root transformation for the odds ratio can be applied so as to approximate the risk ratio.\textsuperscript{24} In this way all the parameters that the multi-bias E-value pertains to are on the (approximate) risk ratio scale.

Figure 2 shows the size of the multiple-bias E-value for various combinations of biases and across a range of observed risk ratios. In general, this demonstrates that when there are multiple forms of bias, very little of each type could be sufficient to produce a risk ratio that is within the range we generally see in epidemiologic studies. For example, when the null is true, it is possible to observe a risk ratio of 4 if each of the outcome misclassification (\(\RR{AY^* \mid y, S = 1}\)), selection bias (\(\RR{U_sY\mid A = 1}, \RR{U_sY\mid A = 1}, \RR{SU_s\mid A = 1}, \RR{SU_s\mid A = 0}\)), and unmeasured confounding (\(\RR{U_cY }, \RR{AU_c}\)) parameters is approximately 1.89.

Of course, it is unlikely that each of these sensitivity analysis parameters would be equal to the others, and equal to 1.89. The bounds in this article can be used to assess the bias with a more realistic set of parameters. However, comparing multiple-bias E-values for various combinations of biases may be useful when planning studies to assess where resources should be invested to avoid certain biases, or to assess where a more in-depth bias analysis would be most useful.

Unfortunately, we know of no closed-form solution for this value when we are faced with all three types of bias, but it is easily solved numerically. The expressions to be solved are given in the final column of Table 1. To calculate the analogous multi-bias E-value needed to shift the observed \(\text{RR}^{\text{obs}}_{AY}\) to some risk ratio, \(\text{RR}^{\text{true}}_{AY}\), other than the null, one can simply replace \(\text{RR}^{\text{obs}}_{AY}\) in the each formula with \(\text{RR}^{\text{obs}}_{AY}/ \text{RR}^{\text{true}}_{AY}\). Also, each formula presupposes that \(\text{RR}^{\text{obs}}_{AY}>= 1\); for apparently protective exposures, the inverse should be taken first.

We will demonstrate interpretation of the multiple bias E-value with respect to our examples, and then briefly describe an R package that can be used to implement the results.

\hypertarget{examples}{%
\subsubsection{Examples}\label{examples}}

Recall that the study of HIV infection in children found \(\text{RR}^{\text{obs}}_{AY}= 6.75\),\textsuperscript{30} which we determined was possibly affected by selection bias and unmeasured confounding. The multi-bias E-value for that study, given the assumptions about bias we have made, is 4.64. This tells us that \(\RR{U_sY \mid A = 1} = \RR{SU_s \mid A = 1} = \RR{AU_c} = \RR{U_cY} = \geq 4.64\) could suffice to completely explain the observed result, but weaker combined bias would not. If, for example, selection bias were indeed weaker, the strength of the unmeasured confounding parameters would have to be stronger than 4.64 for the observation to be compatible with a truly null effect. Repeating the calculation with the lower limit of the confidence interval, we obtain a multi-bias E-value of 2.73. If all of the parameters were this large, it is possible that the confidence interval would include the null.

The estimate from the vitamins-leukemia study was \(\text{RR}^{\text{obs}}_{A^*Y}= 0.51\).\textsuperscript{31} After taking the inverse so that\(\text{RR}^{\text{obs}}_{A^*Y}= 1/0.51 = 1.96\), we find that the multi-bias E-value for exposure misclassification and unmeasured confounding is 1.35. In order to interpret that number consistently across biases, the multi-bias E-value we have calculated pertains to \(\RR{AU_c}\), \(\RR{U_cY}\), and \(\RR{YA^* \mid a}\), the latter being the square-root approximation of the \(\OR{YA^* \mid a}\) term in the bound for exposure misclassification.\textsuperscript{24} This allows us to interpret 1.35 as the minimum strength on the risk ratio scale that an unmeasured confounder, or set of confounders, would have to have on the outcome, that would have to relate vitamin use to the confounder, and that the false positive probability or sensitivity for vitamin use would have to be increase by, in order for these biases to explain the entire observed risk ratio. Again, this is simply a heuristic, not something we would expect to be the case; for example, we might expect weaker misclassification but stronger confounding. For the limit of the confidence interval closest to the null, 0.89, if we take inverses, we obtain \(1/0.89=1.12\) and the multi-bias E-value for this is only 1.06, indicating that whether the true risk ratio is smaller than or greater than 1 is indeed sensitive to relatively small amounts of bias.

\hypertarget{software}{%
\section*{Software}\label{software}}

The R package EValue\textsuperscript{38} allows for easy calculation of the multiple-bias bounds and the multi-bias E-value for various combinations of biases and assumptions. The new function \texttt{multi\_bias()} creates a set of biases according to the user's specifications, which can then be used with functions \texttt{multi\_bound()} and \texttt{multi\_evalue()} to calculate a bound or a multi-bias E-value.

For example, the biases in the HIV example can be set with \texttt{HIV\_biases\ \textless{}-\ multi\_bias(confounding(),\ selection(\ditto general\ditto , \ditto increased\ risk\ditto ))}. The command to calculate the bound is then \texttt{multi\_bound(biases\ =\ HIV\_biases,\ RRAUc\ =\ 2.3,\ RRUcY\ =\ 2.5,\ RRUsYA1\ =\ 3,\ RRSUsA1\ =\ 2)} and to calculate the E-value it is \texttt{multi\_evalue(biases\ =\ HIV\_biases,\ est\ =\ OR(6.75,\ rare\ =\ TRUE),\ lo\ =\ 2.79,\ hi\ =\ 16.31)}. Similarly, for the vitamins-leukemia example, the biases are set with \texttt{leuk\_biases\ \textless{}-\ multi\_bias(confounding(),\ misclassification(\ditto exposure\ditto ,\ rare\_outcome\ =\ TRUE,\ rare\_exposure\ =\ FALSE))} and the respective bound and E-value commands are \texttt{multi\_bound(biases\ =\ leuk\_biases,\ RRAUc\ =\ 2,\ RRUcY\ =\ 1.22,\ RRYAa\ =\ 1.59)} and \texttt{multi\_evalue(biases\ =\ leuk\_biases,\ est\ =\ OR(0.51,\ rare\ =\ TRUE),\ lo\ =\ 0.30,\ hi\ =\ 0.89)}. More examples are available in the eAppendix, and the package documentation is available online.

\hypertarget{discussion}{%
\section*{Discussion}\label{discussion}}

We have described an approach to sensitivity analysis that we hope can help bridge the gap between complex methods that require specifying many parameters and making restrictive assumptions, and simpler methods that allow for assessment of only one type of bias at a time. The multiple-bias bound can be used to simultaneously consider the possible effects of biases that are of different strengths. Researchers can propose values for the parameters based on background knowledge, validation studies, or simply hypothetical situations, and assess the minimum possible true risk ratio that would be compatible if the observed value were affected by biases of that magnitude. When planning for future research, the bound can be used to compare the effects of biases within a given situation and prioritize more extensive confounder assessment, a more valid sampling/inclusion scheme, and/or better measurement techniques. It may also show that certain improvements to study design are futile; if the amount of an unavoidable bias greatly attenuates the anticipated risk ratio estimate, investing resources into reducing another type of bias may not be worth it.\textsuperscript{39,40}

The multi-bias E-value presents an even more straightforward alternative: calculate the joint minimum of the bias parameters that would be required to \enquote{explain away} an observed risk ratio. All of the considerations and caveats of the E-value for unmeasured confounding pertain to this metric as well,\textsuperscript{41} and its more complex interpretation may strain credulity. However, we argue that it can be an interesting thought exercise when evaluating alternative explanations for a result. Even when the multi-bias E-value is used, it can be helpful for researchers to also employ the multiple-bias bounds for their best-informed assessments of the sensitivity parameters, and additionally to consider multi-bias E-values not only for the null but also for minimal effect magnitudes that would be considered of scientific or public health importance.

Though the calculations involved in our approach are simple, the entire process of assessing bias should not be. Importantly, it should be specific to the study design, the available data, and the research question; values for the sensitivity parameters and the multi-bias E-value are meaningless without a frame of reference. Unmeasured confounders could be anything from a single missing risk factor to the \enquote{ultimate covariate,}\textsuperscript{42} the variable encoding an individual's causal type. Misclassification may be negligible or close to non-differential, or as bad as chance in one or another group; it is up to researchers and readers to assess the plausibility of these situations with respect to a given study and what was conditioned on in the analysis, and then assess how much bias they would create.

While we have suggested two possible orderings for factoring the bias, others that take into account, for example, misclassification that is also differential by an unmeasured confounder, are possible. We have presented results for risk ratios, which can in many cases be extended to odds ratios. Further work would need to be done to extend this approach to risk differences or mean differences, which may be especially challenging because the bounds are more frequently non-informative.\textsuperscript{43} Other approaches exist to quantify as simply as possible unmeasured confounding in linear or probit models,\textsuperscript{44--47} but to our knowledge they have not yet been extended to multiple biases.

There is no single solution to the problem of bias in epidemiologic research. Some biases can be corrected at the design phase, others in the main analysis, but the assessment of what bias may remain should be a regular component of any study that attempts to quantify causal effects. Here we have presented one method that can make it simpler to do so.

\newpage
\begin{singlespace}Table  1: Multiple bias bounds for various combinations of biases. The first three columns show different combinations and ordering of the biases, as well as the implied assumptions. The fourth column contains the expression that bounds the bias when those assumptions hold. The definitions of the parameters are given in the main text. The final column contains the expressions that can be solved to calculate the multi-bias E-value, which refers to the parameters in the previous column. If a non-null multi-bias E-value is desired, one can simply replace $\text{RR}^{\text{obs}}_{AY}$ in the formulae with $\text{RR}^{\text{obs}}_{AY}/ \text{RR}^{\text{true}}_{AY}$.\end{singlespace}

\newgeometry{left=.15cm,bottom=1cm,right=.15cm,top=1cm}
\begin{landscape}
\begin{singlespace}
\begin{longtable}{P{.23\textwidth}P{.23\textwidth}P{.23\textwidth}P{.4\textwidth}P{.17\textwidth}}
\toprule

\multicolumn{3}{c}{Biases and associated assumptions} &
  Bound &
  E-value solution \\ \midrule

unmeasured confounding \newline $Y_a \amalg A \mid C, U_c$
  &  general selection bias \newline $Y\amalg S \mid A, C, U_s$  
  & outcome misclassification 
  & $g\left(\RR{AU_c}, \RR{U_cY}\right) \times
  g\left(\RR{U_sY \mid A = 1}, \RR{SU_s \mid A = 1}\right) \times g\left(\RR{U_sY \mid A = 0}\RR{SU_s \mid A = 0}\right) \times \RR{AY^* \mid y, S = 1}$ 
  &  $\frac{x^7}{(2x-1)^3} \geq \text{RR}^{\text{obs}}_{AY^*}$ \\
  
  \\

unmeasured confounding \newline $Y_a \amalg A \mid C, U_c$
  & outcome misclassification 
    & general selection bias \newline $Y^*\amalg S \mid A, C, U_s$ 
  & $g\left(\RR{AU_c}, \RR{U_cY}\right) \times
  \RR{AY^* \mid y} \times 
  g\left(\RR{U_sY^* \mid A = 1}, \RR{SU_s \mid A = 1}\right) \times g\left(\RR{U_sY^* \mid A = 0}\RR{SU_s \mid A = 0}\right)$ 
  &  $\frac{x^7}{(2x-1)^3} \geq \text{RR}^{\text{obs}}_{AY^*}$ \\
   
  \\
  \\

unmeasured confounding 
  & selected population 
  & outcome misclassification 
  & $g\left(\RR{AU_{sc}\mid S = 1}, \RR{U_{sc}Y\mid S = 1}\right) \times
  \RR{AY^* \mid y} $ 
  &  $\frac{x^3}{(2x-1)} \geq \text{RR}^{\text{obs}}_{AY^*}$ \\

\multicolumn{2}{c}{$Y_a \amalg A \mid S = 1, C, U_c, U_s \;\;\;\;\;\;\;$} \\

  \\
  \\
unmeasured confounding \newline $Y_a \amalg A \mid C, U_c$
  &  general selection bias \newline $Y\amalg S \mid A, C, U_s$  
  & exposure misclassification \newline $\Prc{Y = 0}{a, c, S = 1} \approx 1$
  & $g\left(\RR{AU_c}, \RR{U_cY}\right) \times
  g\left(\RR{U_sY \mid A = 1}, \RR{SU_s \mid A = 1}\right) \times g\left(\RR{U_sY \mid A = 0}\RR{SU_s \mid A = 0}\right) \times 
  \OR{YA^* \mid a, S = 1}$ 
  &  $\frac{x^8}{(2x-1)^3} \geq \text{RR}^{\text{obs}}_{A^*Y}$ \\
  \\

 unmeasured confounding \newline $Y_a \amalg A \mid C, U_c$
  & exposure misclassification \newline 
  $\Prc{Y = 0}{a, c} \approx 1$
  &  general selection bias \newline $Y\amalg S \mid A^*, C, U_s$ 
  & $g\left(\RR{AU_c}, \RR{U_cY}\right) \times
  \OR{YA^* \mid a} \times
  g\left(\RR{U_sY \mid A^* = 1}, \RR{SU_s \mid A^* = 1}\right) \times g\left(\RR{U_sY \mid A^* = 0}\RR{SU_s \mid A^* = 0}\right)$ 
  &  $\frac{x^8}{(2x-1)^3} \geq \text{RR}^{\text{obs}}_{A^*Y}$ \\
  
  \\
  \\

unmeasured confounding 
  & selected population 
  & exposure misclassification 
  & $g\left(\RR{AU_{sc}\mid S = 1}, \RR{U_{sc}Y\mid S = 1}\right) \times 
  \OR{YA^* \mid a}$ 
  &  $\frac{x^4}{(2x-1)} \geq \text{RR}^{\text{obs}}_{A^*Y}$ \\

\multicolumn{2}{c}{$Y_a \amalg A \mid S = 1, C, U_c, U_s \;\;\;\;\;\;\;$} 
  & $\Prc{Y = 0}{a, c, S = 1} \approx 1$\\
\\
\\

unmeasured confounding \newline $Y_a \amalg A \mid C, U_c$
  &  general selection bias \newline $Y\amalg S \mid A, C, U_s$  
  & exposure misclassification \newline $\Prc{Y = 0}{a, c, S = 1} \approx 1$\newline $\Prc{A = 0}{y, c, S = 1} \approx 1$
  & $g\left(\RR{AU_c}, \RR{U_cY}\right) \times
  g\left(\RR{U_sY \mid A = 1}, \RR{SU_s \mid A = 1}\right) \times g\left(\RR{U_sY \mid A = 0}\RR{SU_s \mid A = 0}\right) \times 
  \RR{YA^* \mid a, S = 1} $ 
  &  $\frac{x^7}{(2x-1)^3} \geq \text{RR}^{\text{obs}}_{A^*Y}$ \\
  \\

unmeasured confounding \newline $Y_a \amalg A \mid C, U_c$
  & exposure misclassification \newline $\Prc{Y = 0}{a,  c} \approx 1$\newline $\Prc{A = 0}{y, c} \approx 1$
  &  general selection bias \newline $Y\amalg S \mid A^*, C, U_s$ 
  & $g\left(\RR{AU_c}, \RR{U_cY}\right) \times
  \RR{YA^* \mid a} \times
  g\left(\RR{U_sY \mid A^* = 1}, \RR{SU_s \mid A^* = 1}\right) \times g\left(\RR{U_sY \mid A^* = 0}\RR{SU_s \mid A^* = 0}\right) $ 
  & $\frac{x^7}{(2x-1)^3} \geq \text{RR}^{\text{obs}}_{A^*Y}$ \\
  
  \\
  \\

unmeasured confounding 
  & selected population 
  & exposure misclassification 
  & $g\left(\RR{AU_{sc}\mid S = 1}, \RR{U_{sc}Y\mid S = 1}\right) \times
  \RR{YA^* \mid a}$ 
  &  $\frac{x^3}{(2x-1)} \geq \text{RR}^{\text{obs}}_{A^*Y}$ \\

\multicolumn{2}{c}{$Y_a \amalg A \mid S = 1, C, U_c, U_s \;\;\;\;\;\;\;$}
  & $\Prc{Y = 0}{a, c, S = 1} \approx 1$ \newline $\Prc{A = 0}{y, c, S = 1} \approx 1$ \\

\bottomrule
\end{longtable}
\end{singlespace}
\end{landscape}
\restoregeometry

\newpage

\begin{singlespace}Figure  1: Directed acyclic graphs depicting the examples described in the text.\end{singlespace}

\begin{center}\includegraphics{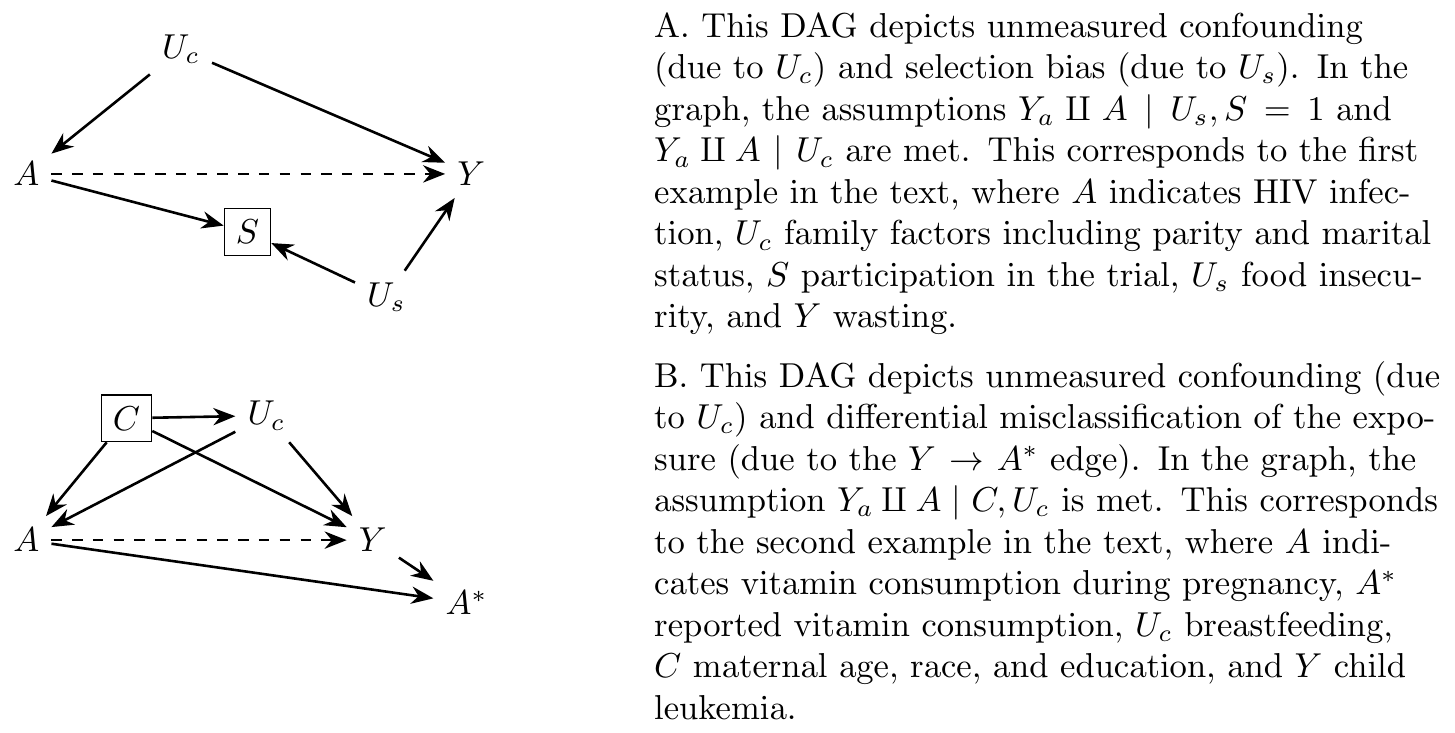} \end{center}

\newpage
\begin{singlespace}Figure  2: Multi-bias E-values for various combinations of biases and for observed risk ratios ranging from 1 to 7.\end{singlespace}

\begin{center}\includegraphics{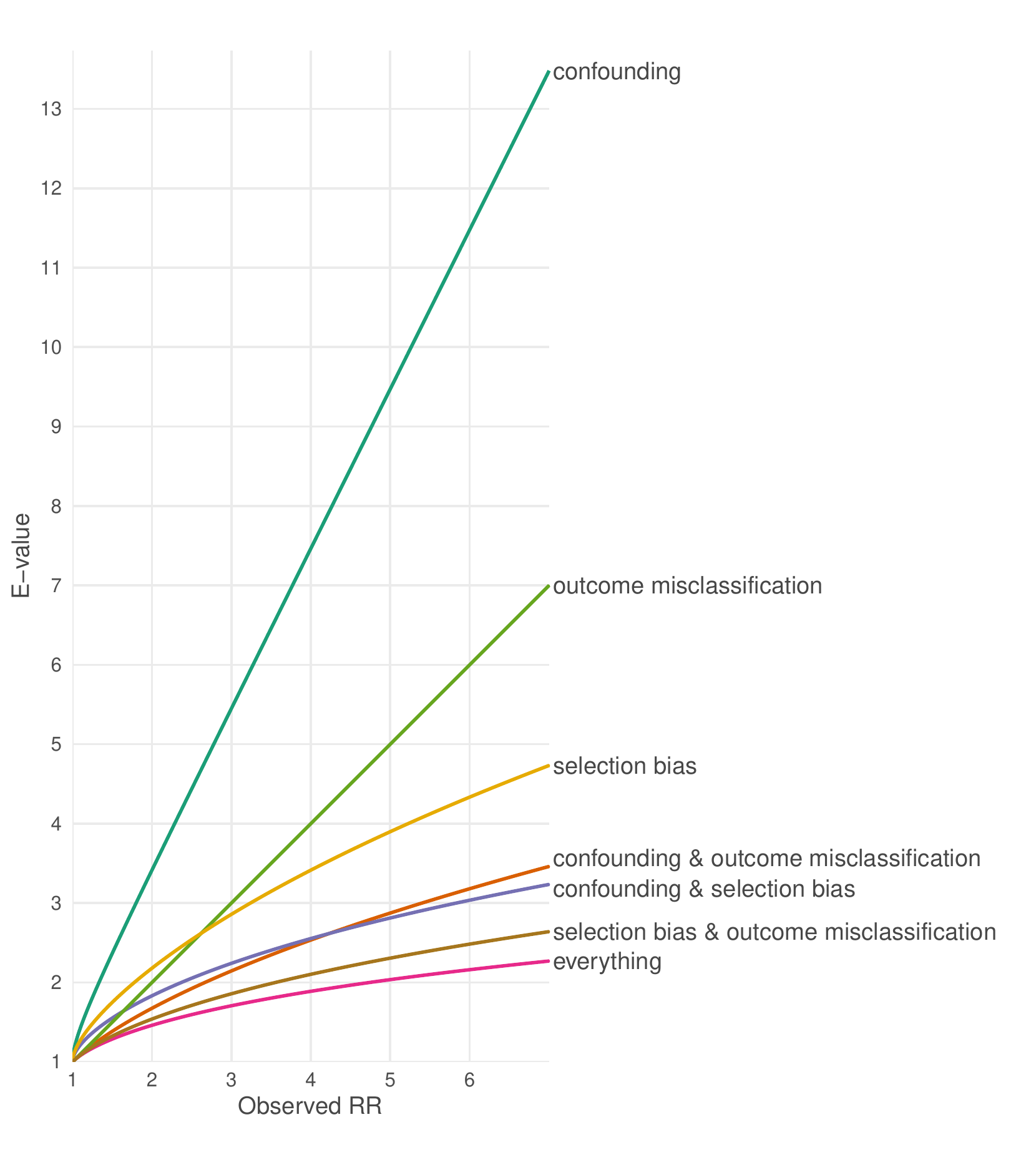} \end{center}

\newpage

\hypertarget{references}{%
\section*{References}\label{references}}

\hypertarget{refs}{}
\leavevmode\hypertarget{ref-Ioannidis2007}{}%
1. Ioannidis JP. Limitations are not properly acknowledged in the scientific literature. \emph{J Clin Epidemiol}. 2007;60:324--329.

\leavevmode\hypertarget{ref-Lash2007}{}%
2. Lash TL. Heuristic thinking and inference from observational epidemiology. \emph{Epidemiology}. 2007;18:67--72.

\leavevmode\hypertarget{ref-Cornfield1959}{}%
3. Cornfield J, Haenszel W, Hammond EC, et al. Smoking and lung cancer: Recent evidence and a discussion of some questions. \emph{J Natl Cancer I}. 1959;22:173--203.

\leavevmode\hypertarget{ref-Bross1966}{}%
4. Bross IDJ. Spurious effects from an extraneous variable. \emph{J Chron Dis}. 1966;19:637--647.

\leavevmode\hypertarget{ref-Bross1967}{}%
5. Bross IDJ. Pertinency of an extraneous variable. \emph{J Chron Dis}. 1967;20:487--495.

\leavevmode\hypertarget{ref-Schlesselman1978}{}%
6. Schlesselman JJ. Assessing effects of confounding variables. \emph{Am J Epidemiol}. 1978;108:3--8.

\leavevmode\hypertarget{ref-Rosenbaum1983}{}%
7. Rosenbaum PR, Rubin DB. The central role of the propensity score in observational studies for causal effects. \emph{Biometrika}. 1983;70:41--55.

\leavevmode\hypertarget{ref-Copeland1977}{}%
8. Copeland KT, Checkoway H, McMichael AJ, et al. Bias due to misclassification in the estimation of relative risk. \emph{Am J Epidemiol}. 1977;105:488--495.

\leavevmode\hypertarget{ref-Barron1977}{}%
9. Barron BA. The effects of misclassification on the estimation of relative risk. \emph{Biometrics}. 1977;33:414--418.

\leavevmode\hypertarget{ref-Greenland1981}{}%
10. Greenland S, Neutra R. An analysis of detection bias and proposed corrections in the study of estrogens and endometrial cancer. \emph{J Chron Dis}. 1981;34:433--438.

\leavevmode\hypertarget{ref-Greenland1983}{}%
11. Greenland S, Ericson C, Kleinbaum DG. Correcting for misclassification in two-way tables and matched-pair studies. \emph{Int J Epidemiol}. 1983;12:93--97.

\leavevmode\hypertarget{ref-Lash2000}{}%
12. Lash TL, Silliman RA. A sensitivity analysis to separate bias due to confounding from bias due to predicting misclassification by a variable that does both. \emph{Epidemiology}. 2000;11:544--549.

\leavevmode\hypertarget{ref-Greenland2003a}{}%
13. Greenland S. The impact of prior distributions for uncontrolled confounding and response bias: A case study of the relation of wire codes and magnetic fields to childhood leukemia. \emph{J Am Stat Assoc}. 2003;98:47--54.

\leavevmode\hypertarget{ref-Lash2003}{}%
14. Lash TL, Fink AK. Semi-automated sensitivity analysis to assess systematic errors in observational data. \emph{Epidemiology}. 2003;14:451--458.

\leavevmode\hypertarget{ref-Fox2005a}{}%
15. Fox MP, Lash TL, Greenland S. A method to automate probabilistic sensitivity analyses of misclassified binary variables. \emph{Int J Epidemiol}. 2005;34:1370--1376.

\leavevmode\hypertarget{ref-Greenland2005}{}%
16. Greenland S. Multiple-bias modelling for analysis of observational data (with discussion). \emph{J Roy Stat Soc A}. 2005;168:267--306.

\leavevmode\hypertarget{ref-Lash2010}{}%
17. Lash TL, Schmidt M, Jensen AØ, et al. Methods to apply probabilistic bias analysis to summary estimates of association. \emph{Pharmacoepidem Dr S}. 2010;19:638--644.

\leavevmode\hypertarget{ref-Lash2009}{}%
18. Lash TL, Fox MP, Fink AK. Applying quantitative bias analysis to epidemiologic data.; 2009. doi:\href{https://doi.org/10.1111/j.1467-985x.2010.00681_10.x}{10.1111/j.1467-985x.2010.00681\_10.x}.

\leavevmode\hypertarget{ref-Orsini2008}{}%
19. Orsini N, Bellocco R, Bottai M, et al. A tool for deterministic and probabilistic sensitivity analysis of epidemiologic studies. \emph{Stata J}. 2008;8:29--48.

\leavevmode\hypertarget{ref-Hunnicutt2016}{}%
20. Hunnicutt JN, Ulbricht CM, Chrysanthopoulou SA, et al. Probabilistic bias analysis in pharmacoepidemiology and comparative effectiveness research: A systematic review. \emph{Pharmacoepidem Dr S}. 2016;25:1343--1353.

\leavevmode\hypertarget{ref-Lash2014a}{}%
21. Lash TL, Abrams B, Bodnar LM. Comparison of bias analysis strategies applied to a large data set. \emph{Epidemiology}. 2014;25:576--582.

\leavevmode\hypertarget{ref-Ding2016b}{}%
22. Ding P, VanderWeele TJ. Sensitivity analysis without assumptions. \emph{Epidemiology}. 2016;27:368--377.

\leavevmode\hypertarget{ref-Smith2019}{}%
23. Smith LH, VanderWeele TJ. Bounding bias due to selection. \emph{Epidemiology}. 2019;30:509--516.

\leavevmode\hypertarget{ref-VanderWeele2019b}{}%
24. VanderWeele TJ, Li Y. Simple sensitivity analysis for differential measurement error. \emph{Am J Epidemiol}. 2019;188:1823--1829.

\leavevmode\hypertarget{ref-VanderWeele2017a}{}%
25. VanderWeele TJ, Ding P. Sensitivity analysis in observational research: Introducing the e-value. \emph{Ann Intern Med}. 2017;167:268--275.

\leavevmode\hypertarget{ref-Blum2020}{}%
26. Blum MR, Tan YJ, Ioannidis JPA. Use of e-values for addressing confounding in observational studies-an empirical assessment of the literature. \emph{Int J Epidemiol}. 2020;1--13.

\leavevmode\hypertarget{ref-Chin2017}{}%
27. Chin HB, Baird DD, McConnaughey DR, et al. Long-term recall of pregnancy-related events. \emph{Epidemiology}. 2017;28:575--579.

\leavevmode\hypertarget{ref-Greene2011}{}%
28. Greene N, Greenland S, Olsen JJ, et al. Estimating bias from loss to follow-up in the danish national birth cohort. \emph{Epidemiology}. 2011;22:1.

\leavevmode\hypertarget{ref-Mumford2018}{}%
29. Mumford SL, Yeung EH. Intergenerational effects--causation or confounding? \emph{Fertil Steril}. 2018;110:52--53.

\leavevmode\hypertarget{ref-Omoni2017}{}%
30. Omoni AO, Ntozini R, Evans C, et al. Child growth according to maternal and child hiv status in zimbabwe. \emph{Pediatr Infect Dis J}. 2017;36:869--876.

\leavevmode\hypertarget{ref-Ross2005}{}%
31. Ross JA, Blair CK, Olshan AF, et al. Periconceptional vitamin use and leukemia risk in children with down syndrome: A children's oncology group study. \emph{Ann Ny Acad Sci}. 2005;104:405--410.

\leavevmode\hypertarget{ref-Phillips2003a}{}%
32. Phillips CV. Quantifying and reporting uncertainty from systematic errors. \emph{Epidemiology}. 2003;14:459--466.

\leavevmode\hypertarget{ref-Greenland1996a}{}%
33. Greenland S. Basic methods for sensitivity analysis of biases. \emph{Int J Epidemiol}. 1996;25:1107--1116.

\leavevmode\hypertarget{ref-Maclure2001}{}%
34. Maclure M, Schneeweiss S. Causation of bias: The episcope. \emph{Epidemiology}. 2001;12:114--122.

\leavevmode\hypertarget{ref-Jurek2009}{}%
35. Jurek AM, Maldonado G, Spector LG, et al. Periconceptional maternal vitamin supplementation and childhood leukaemia: An uncertainty analysis. \emph{J Epidemiol Commun H}. 2009;63:168--172.

\leavevmode\hypertarget{ref-Amitay2015}{}%
36. Amitay EL, Keinan-Boker L. Breastfeeding and childhood leukemia incidence: A meta-analysis and systematic review. \emph{JAMA Pediatrics}. 2015;169:1--9.

\leavevmode\hypertarget{ref-Phillips2003b}{}%
37. Phillips CV, LaPole LM. Quantifying errors without random sampling. \emph{BMC Medical Research Methodology}. 2003;3:1--10.

\leavevmode\hypertarget{ref-Mathur2018}{}%
38. Mathur MB, Ding P, Riddell CA, et al. Website and r package for computing e-values. \emph{Epidemiology}. 2018;29:e45--e47.

\leavevmode\hypertarget{ref-Lash2012}{}%
39. Lash TL, Ahern TP. Bias analysis to guide new data collection. \emph{Int J Biostat}. 2012;8. doi:\href{https://doi.org/10.2202/1557-4679.1345}{10.2202/1557-4679.1345}.

\leavevmode\hypertarget{ref-Fox2020}{}%
40. Fox MP, Lash TL. Quantitative bias analysis for study and grant planning. \emph{Ann Epidemiol}. 2020;43:32--36.

\leavevmode\hypertarget{ref-VanderWeele2019a}{}%
41. VanderWeele TJ, Ding P, Mathur M. Technical considerations in the use of the e-value. \emph{Journal of Causal Inference}. 2019;7. doi:\href{https://doi.org/10.1515/jci-2018-0007}{10.1515/jci-2018-0007}.

\leavevmode\hypertarget{ref-Greenland1986}{}%
42. Greenland S, Robins JM. Identifiability, exchangeability, and epidemiological confounding. \emph{Int J Epidemiol}. 1986;15:413--419.

\leavevmode\hypertarget{ref-Ding2014}{}%
43. Ding P, Vanderweele TJ. Generalized cornfield conditions for the risk difference. \emph{Biometrika}. 2014;101:971--977.

\leavevmode\hypertarget{ref-Frank2000}{}%
44. Frank KA. Impact of a confounding variable on a regression coefficient. \emph{Sociological Methods and Research}. 2000;29:147--194.

\leavevmode\hypertarget{ref-Altonji2005}{}%
45. Altonji JG, Elder TE, Taber CR. Selection on observed and unobserved variables: Assessing the effectiveness of catholic schools. \emph{J Polit Econ}. 2005;113:151--184.

\leavevmode\hypertarget{ref-Oster2019}{}%
46. Oster E. Unobservable selection and coefficient stability: Theory and evidence. \emph{Journal of Business and Economic Statistics}. 2019;37:187--204.

\leavevmode\hypertarget{ref-Cinelli2019}{}%
47. Cinelli C, Hazlett C. Making sense of sensitivity: Extending omitted variable bias. \emph{J Roy Stat Soc B}. 2019;39--67.

\clearpage

\pagenumbering{gobble} 

\begin{center}
  \vspace{-4ex}
  \TitleFont{
  eAppendix
  }

\end{center}

{
\setcounter{tocdepth}{2}
\tableofcontents
\addtocontents{toc}{\protect\thispagestyle{empty}}
\setcounter{page}{0}
}
\newpage
\pagenumbering{arabic}

\hypertarget{a-bound-for-outcome-misclassification-selection-bias-and-unmeasured-confounding}{%
\section{A bound for outcome misclassification, selection bias, and unmeasured confounding}\label{a-bound-for-outcome-misclassification-selection-bias-and-unmeasured-confounding}}

\hypertarget{result-1}{%
\subsection{Result 1}\label{result-1}}

Let \(A\) denote a binary exposure of interest, \(Y\) a binary outcome and \(Y^*\) the misclassified version, and \(C\) measured covariates. Additionally let \(S\) be a binary indicator of selection into a study, so that we can collect data only on the subset of the population for which \(S = 1\). Finally, assume that there exist \(U_s\) and \(U_c\) such that \(Y \amalg S \mid A, C, U_s\) and \(Y_a \amalg A \mid C, U_c\), but that it is not necessarily true that \(Y \amalg S \mid A, C\) or \(Y_a \amalg A \mid C\).

We can estimate a confounded risk ratio observed in the selected population, subject to (potentially differential) outcome misclassification, \(\text{RR}^{\text{obs}}_{AY}\), but our inferential goal is a causal risk ratio for the true outcome in the entire population, \(\text{RR}^{\text{true}}_{AY}\):
\begin{align*}
\text{RR}^{\text{obs}}_{AY}& = \frac{\Prc{Y^* = 1}{A = 1, S = 1, c}}{\Prc{Y^* = 1}{A = 0, S = 1, c}} \\
\text{RR}^{\text{true}}_{AY}& = \frac{\Prc{Y_1 = 1}{c}}{\Prc{Y_0 = 1}{c}}
\end{align*}

We have from VanderWeele \& Li\textsuperscript{1} that, for \(\text{RR}^{\text{true}}_{AY}\geq 1\),
\begin{align}
\label{eq:mis}
\text{RR}^{\text{obs}}_{AY}\leq  \text{BF}_{m}\times \frac{\Prc{Y = 1}{A = 1, S = 1, c}}{\Prc{Y = 1}{A = 0, S = 1, c}} 
\end{align}
for
\begin{align}
\label{bf:misorig}
\text{BF}_{m}= \RR{AY^* \mid y, S = 1} = \max_y{\frac{\Prc{Y^* = 1}{Y = y, A = 1, S = 1, c}}{\Prc{Y^* = 1}{Y = y, A = 0, S = 1, c}}} \;.
\end{align}

Then, since we are assuming that \(Y \amalg S \mid A, C, U_s\), from Smith \& VanderWeele\textsuperscript{2} we have that
\begin{align}
\label{eq:sel}
\frac{\Prc{Y = 1}{A = 1, S = 1, c}}{\Prc{Y = 1}{A = 0, S = 1, c}} \leq \text{BF}_{s}\times \frac{\Prc{Y = 1}{A = 1, c}}{\Prc{Y = 1}{A = 0, c}}
\end{align}
for
\begin{align*}
\text{BF}_{s}= \frac{\RR{U_sY\mid A = 1} \times \RR{SU_s\mid A = 1}}{\RR{U_sY\mid A = 1} + \RR{SU_s\mid A = 1} - 1} \times \frac{\RR{U_sY\mid A = 0} \times \RR{SU_s\mid A = 0}}{\RR{U_sY\mid A = 0} + \RR{SU_s\mid A = 0} - 1}
\end{align*}
where
\begin{align}
\RR{U_sY\mid A = a} & = \frac{\max_u \Prc{Y = 1}{A = a, c, U_s = u}}{\min_u \Prc{Y = 1}{A = a, c, U_s = u}}\;\; \text{for } a = 0, 1 \nonumber \\
\RR{SU_s\mid A = 1} & = \max_u\frac{\Prc{U_s = u}{A = 1, S = 1, c}}{\Prc{U_s = u}{A = 1, S = 0, c}} \nonumber \\
\RR{SU_s\mid A = 0} & = \max_u\frac{\Prc{U_s = u}{A = 0, S = 0, c}}{\Prc{U_s = u}{A = 0, S = 1, c}}\;. \label{bf:selorig}
\end{align}

Finally, since we are assuming that \(Y_a \amalg A \mid C, U_c\) from Ding \& VanderWeele\textsuperscript{3} we have
\begin{align}
\label{eq:con}
\frac{\Prc{Y = 1}{A = 1, c}}{\Prc{Y = 1}{A = 0, c}} \leq \text{BF}_{c}\times \frac{\Prc{Y_1 = 1}{c}}{\Prc{Y_0 = 1}{c}}
\end{align}
for
\begin{align}
\text{BF}_{c}= \frac{\RR{AU_c} \times \RR{U_cY}}{\RR{AU_c} + \RR{U_cY} - 1} \label{bf:con}
\end{align}
where
\begin{align*}
\RR{AU_c} & = \max_u \frac{\Prc{U_c = u}{A = 1, c}}{\Prc{U_c = u}{A = 0, c}} \\
\RR{U_cY } & = \max_a\frac{\max_u \Prc{Y = 1}{A = a, c, U_c = u}}{\min_u \Prc{Y = 1}{A = a, c, U_c = u}}\;.
\end{align*}

Putting together expressions \eqref{eq:mis}, \eqref{eq:sel}, and \eqref{eq:con}, we have Result 1:
\begin{align}
\text{RR}^{\text{obs}}_{AY}& \leq \text{BF}_{m}\times \frac{\Prc{Y = 1}{A = 1, S = 1, c}}{\Prc{Y = 1}{A = 0, S = 1, c}}  \nonumber \\
& \leq \text{BF}_{m}\times \text{BF}_{s}\times \frac{\Prc{Y = 1}{A = 1, c}}{\Prc{Y = 1}{A = 0, c}} \nonumber \\
& \leq  \text{BF}_{m}\times \text{BF}_{s}\times \text{BF}_{c}\times \frac{\Prc{Y_1 = 1}{c}}{\Prc{Y_0 = 1}{c}} \nonumber \\
& = \text{BF}_{m}\times \text{BF}_{s}\times \text{BF}_{c}\times \text{RR}^{\text{true}}_{AY}\;. \label{eq:all}
\end{align}

\hypertarget{an-alternative-decomposition}{%
\subsection{An alternative decomposition}\label{an-alternative-decomposition}}

Now assume that there exist \(U_s\) and \(U_c\) such that \(Y^* \amalg S \mid A, C, U_s\) and \(Y_a \amalg A \mid C, U_c\). This may be the case if, for example, selection into the study is based on a factor related to the (mis)measured outcome, not the true outcome.

Then we can bound the bias with the same final expression, but some of the parameters within the bias factors are defined slightly differently.

The possible magnitude of selection bias can be defined in terms of the misclassified outcome, so that
\begin{align*}
\text{BF}_{s}= \frac{\RR{U_sY^* \mid A = 1} \times \RR{SU_s\mid A = 1}}{\RR{U_sY^* \mid A = 1} + \RR{SU_s \mid A = 1} - 1} \times \frac{\RR{U_sY^* \mid A = 0} \times \RR{SU_s\mid A = 0}}{\RR{U_sY^*\mid A = 0} + \RR{SU_s\mid A = 0} - 1}
\end{align*}
where
\begin{align*}
\RR{U_sY^* \mid A = a} & = \frac{\max_u \Prc{Y^* = 1}{A = a, c, U_s = u}}{\min_u \Prc{Y^* = 1}{A = a, c, U_s = u}}\;\; \text{for } a = 0, 1
\end{align*}
and \(\RR{SU_s\mid A = 1}\) and \(\RR{SU_s\mid A = 0}\) are defined as in \eqref{bf:selorig} above.

Then, the measurement error correction applies to the entire population, so that
\begin{align*}
\text{BF}_{m}= \RR{AY^* \mid y} = \max_y{\frac{\Prc{Y^* = 1}{Y = y, A = 1, c}}{\Prc{Y^* = 1}{Y = y, A = 0, c}}} \;.\end{align*}

Expression \eqref{eq:all} now holds with the newly defined \(\text{BF}_{s}\) and \(\text{BF}_{m}\).

\hypertarget{a-bound-for-exposure-misclassification-selection-bias-and-unmeasured-confounding}{%
\section{A bound for exposure misclassification, selection bias, and unmeasured confounding}\label{a-bound-for-exposure-misclassification-selection-bias-and-unmeasured-confounding}}

Unlike the bound for outcome misclassification, the bound for exposure misclassification from VanderWeele \& Li\textsuperscript{1} applies to the odds ratio, not the risk ratio, and the sensitivity parameters are also not risk ratios. That is,
\begin{align}
\label{eq:misex}
\frac{\nicefrac{\Prc{Y = 1}{A^* = 1, c}}{\Prc{Y = 0}{A^* = 1, c}}}{\nicefrac{\Prc{Y = 1}{A^* = 0, c}}{\Prc{Y = 0}{A^* = 0, c}}} \leq  \text{BF}_{m}' \times \frac{\nicefrac{\Prc{Y = 1}{A = 1, c}}{\Prc{Y = 0}{A = 1, c}}}{\nicefrac{\Prc{Y = 1}{A = 0, c}}{\Prc{Y = 0}{A = 0, c}}}
\end{align}
for
\begin{align}
\text{BF}_{m}' = \text{OR}_{YA^*\mid a} = \max\left(\frac{\nicefrac{s'_1}{1 - s'_1}}{\nicefrac{s'_0}{1 - s'_0}}, \frac{\nicefrac{f'_1}{1 - f'_1}}{\nicefrac{f'_0}{1 - f'_0}}, \frac{\nicefrac{f'_1}{ f'_0}}{\nicefrac{1 - s'_1}{1 - s'_0}}, \frac{\nicefrac{s'_1}{s'_0}}{\nicefrac{1 - f'_1}{1 - f'_0}}\right) \label{eq:bmprime}
\end{align}
where \(s'_y = \Prc{A^* = 1}{Y = y, A = 1, c}\) and \(f'_y = \Prc{A^* = 1}{Y = y, A = 0, c}\).

Applying this bound after factoring out selection bias, we would find that we are left with
\begin{align*}
\text{RR}^{\text{obs}}_{AY}\leq \text{BF}_{m}' \times \text{BF}_{s}\times \text{BF}_{c}\times \text{RR}^{\text{true}}_{AY}\times \frac{\Prc{Y = 0}{A = 0, c}}{\Prc{Y = 0}{A = 1, c}} \times \frac{\Prc{Y = 0}{A^* = 1, c}}{\Prc{Y = 0}{A^* = 0, c}}
\end{align*}
for some \(\text{BF}_{m}'\), \(\text{BF}_{s}\), and \(\text{BF}_{c}\), which is not as useful for sensitivity analysis. However, if the outcome is sufficiently rare that \(\Prc{Y = 0}{\cdot} \approx 1\) in all strata, a simpler bound holds approximately, as we show next.

Again we can define the parameters in the bound in two ways by considering two sets of assumptions.

\hypertarget{result-2}{%
\subsection{Result 2}\label{result-2}}

If there exist \(U_s\) and \(U_c\) such that \(Y \amalg S \mid A, C, U_s\) and \(Y_a \amalg A \mid C, U_c\), and if \(\Prc{Y = 0}{\cdot} \approx 1\), then we have Result 2:
\begin{align*}
\text{RR}^{\text{obs}'}_{AY} & = \frac{\Prc{Y = 1}{A^* = 1, S = 1, c}}{\Prc{Y = 1}{A^* = 0, S = 1, c}} \\
& \lesssim \text{BF}_{m}'\times \text{BF}_{s}\times \text{BF}_{c}\times \text{RR}^{\text{true}}_{AY}
\end{align*}
for \(\text{BF}_{m}' = \text{OR}_{YA^*\mid a, S = 1}\) equivalent to the expression \eqref{eq:bmprime}, but with \(s'_y = \Prc{A^* = 1}{Y = y, A = 1, S = 1, c}\) and \(f'_y = \Prc{A^* = 1}{Y = y, A = 0, S = 1, c}\); \(\text{BF}_{s}\) as defined in \eqref{bf:selorig}; and \(\text{BF}_{c}\) as defined in \eqref{bf:con}.

\hypertarget{an-alternative-decomposition-1}{%
\subsection{An alternative decomposition}\label{an-alternative-decomposition-1}}

Alternatively, if \(Y \amalg S \mid A^*, C, U_s\) and \(Y_a \amalg A \mid C, U_c\), then the bound holds approximately with
\begin{align*}
\text{BF}_{s}& = \frac{\RR{U_sY\mid A^* = 1} \times \RR{SU_s\mid A^* = 1}}{\RR{U_sY\mid A^* = 1} + \RR{SU_s\mid A^* = 1} - 1} \times \frac{\RR{U_sY\mid A^* = 0} \times \RR{SU_s\mid A^* = 0}}{\RR{U_sY\mid A^* = 0} + \RR{SU_s\mid A^* = 0} - 1}
\end{align*}
where \(\RR{U_sY\mid A^* = a}\) and \(\RR{SU_s\mid A^* = 0}\) are defined as above, with all \(A\) replaced with \(A^*\) and \(Y^*\) replaced with \(Y\), and with \(\text{BF}_{m}'\) as originally defined in expression \eqref{eq:bmprime}.

\hypertarget{interpretation-of-the-exposure-misclassification-parameters}{%
\subsection{Interpretation of the exposure misclassification parameters}\label{interpretation-of-the-exposure-misclassification-parameters}}

While all of the sensitivity parameters we have considered thus far are risk ratios, we have seen that those making up the bound for exposure misclassification are not. If, however, the misclassified exposure is sufficiently rare that \(\Prc{A^* = 0}{\cdot} \approx 1\), then we can interpret the sensitivity parameters as risk ratios:
\begin{align*}
\text{BF}_{m}' &  = \RR{YA^* \mid a}  = \max{a}\left(\frac{\Prc{A^* = 1}{Y = 1, A = a, c}}{\Prc{A^* = 1}{Y = 0, A = a, c}}\right) \;\;\;\; \text{or} \\
\text{BF}_{m}' &  = \RR{YA^* \mid a, S = 1}  = \max{a}\left(\frac{\Prc{A^* = 1}{Y = 1, A = a, S = 1, c}}{\Prc{A^* = 1}{Y = 0, A = a, S = 1, c}}\right) \;.
\end{align*}

Alternatively, if the exposure is not particularly rare, we can interpret the sensitivity parameters as squares of the RR equivalents, using the square-root approximation of the odds ratio.\textsuperscript{4}

\hypertarget{inference-in-the-selected-population2}{%
\section{Inference in the selected population}\label{inference-in-the-selected-population2}}

\hypertarget{result-3-under-outcome-misclassification}{%
\subsection{Result 3 (Under outcome misclassification)}\label{result-3-under-outcome-misclassification}}

It may be that our target of inference is the selected population only, so that
\begin{align*}
\text{RR}^{\text{true}}_{AY\mid S = 1}= \frac{\Prc{Y_1 = 1}{S = 1, c}}{\Prc{Y_0 = 1}{S = 1, c}}\;.
\end{align*}
In this case we need that assumption \(Y_a \amalg A \mid S = 1, C, U_c, U_s\;\): we must simultaneously consider both the factor(s) creating selection bias and the factor(s) creating confounding (which may be one and the same). Let \(U_{sc}\) denote the vector \(\left(U_s, U_c\right)\). Then after factoring out the misclassification bias, we have Result 3:
\begin{align}
\text{RR}^{\text{obs}}_{AY}& \leq \text{BF}_{m}\times \frac{\Prc{Y = 1}{A = 1, S = 1, c}}{\Prc{Y = 1}{A = 0, S = 1, c}} \nonumber \\
& \leq \text{BF}_{m}\times \text{BF}_{sc}\times \frac{\Prc{Y_1 = 1}{S = 1, c}}{\Prc{Y_0 = 1}{S = 1, c}} \nonumber \\
& = \text{BF}_{m}\times \text{BF}_{sc}\times \text{RR}^{\text{true}}_{AY\mid S = 1}\label{eq:selected}
\end{align}
for
\begin{align*}
\text{BF}_{sc}= \frac{\RR{AU_{sc}}\times \RR{U_{sc}Y}}{\RR{AU_{sc}} + \RR{U_{sc}Y} - 1}
\end{align*}
where
\begin{align*}
\RR{AU_{sc}} & = \max_{u} \frac{\Prc{U_{sc} = u}{A = 1, S = 1, c}}{\Prc{U_{sc} = u}{A = 0, S = 1, c}} \\
\RR{U_{sc}Y} & = \max_a\frac{\max_{u} \Prc{Y = 1}{A = a, S = 1, c, U_{sc} = u}}{\min_u \Prc{Y = 1}{A = a, S = 1, c , U_{sc} = u}}
\end{align*}
and \(\text{BF}_{m}\) is defined as in \eqref{bf:misorig}.

\hypertarget{under-exposure-misclassification}{%
\subsection{Under exposure misclassification}\label{under-exposure-misclassification}}

Again we consider the bias due to selection and unmeasured confounding jointly. The bound in expression \eqref{eq:selected} holds with \(\text{BF}_{m}'\) constructed with \(s'_y = \Prc{A^* = 1}{Y = y, A = 1, S = 1, c}\) and \(f'_y = \Prc{A^* = 1}{Y = y, A = 0, S = 1, c}\).

\hypertarget{the-multiple-bias-e-value}{%
\section{The multiple bias E-value}\label{the-multiple-bias-e-value}}

To form a multiple bias E-value,\textsuperscript{5} we can set all of the parameters that make up the terms in the bounds equal to each other, then solve for that value to see what magnitude of bias would result in an \(\text{RR}^{\text{obs}}_{AY}\) of at least the value observed, if \(\text{RR}^{\text{true}}_{AY}= 1\).

For example, for the bound for outcome misclassification, general selection bias, and unmeasured confounding:
\begin{align}
\text{RR}^{\text{obs}}_{AY}& \leq \max \RR{AY^* \mid y, S = 1} \times \frac{\RR{U_sY\mid A = 1} \times \RR{SU_s\mid A = 1}}{\RR{U_sY\mid A = 1} + \RR{SU_s\mid A = 1} - 1} \times  \nonumber\\ 
& \;\;\;\; \frac{\RR{U_sY\mid A = 0} \times \RR{SU_s\mid A = 0}}{\RR{U_sY\mid A = 0} + \RR{SU_s\mid A = 0} - 1} \times \frac{\RR{AU_c} \times \RR{U_cY }}{\RR{AU_c} + \RR{U_cY } - 1} \times 1  \nonumber\\
& = x \times \frac{x^2}{2x - 1} \times \frac{x^2}{2x - 1} \times \frac{x^2}{2x - 1} \nonumber \\
& = \frac{x^7}{(2x - 1)^3} \label{eq:poly}
\end{align}
for \(x = \RR{AY^* \mid y, S = 1}= \RR{U_sY\mid A = 1} = \RR{SU_s\mid A = 1} = \RR{U_sY\mid A = 0} = \RR{SU_s\mid A = 0} = \RR{AU_c} = \RR{U_cY}\).

To our knowledge, this polynomial has no closed-form solution. However, we can easily solve it numerically.

For example, if \(\text{RR}^{\text{obs}}_{AY}= 3\), then \(x = 1.71\), meaning that if each of the parameters were at least 1.71, the observed risk ratio could be consistent with a truly null causal risk ratio. If any of the parameters were smaller than 1.71, others would have to be larger if the causal risk ratio were truly null.

We can solve the inequality for any combination of parameters that make up a particular bound in a given situation (e.g., for outcome misclassification and selection bias only, or for exposure misclassification with a rare outcome and unmeasured confounding). When considering exposure misclassification, to calculate a multiple bias E-value, we first must confirm that the outcome is rare. Then, if the misclassified exposure is rare, we can solve equation \eqref{eq:poly} and interpret it with respect to the appropriate parameters; if the exposure is not rare, we can solve
\begin{align*}
\text{RR}^{\text{obs}}_{AY}& \lesssim \RR{YA^* \mid a, S = 1}^2 \times \frac{\RR{U_sY\mid A = 1} \times \RR{SU_s\mid A = 1}}{\RR{U_sY\mid A = 1} + \RR{SU_s\mid A = 1} - 1} \times  \nonumber\\ 
& \;\;\;\; \frac{\RR{U_sY\mid A = 0} \times \RR{SU_s\mid A = 0}}{\RR{U_sY\mid A = 0} + \RR{SU_s\mid A = 0} - 1} \times \frac{\RR{AU_c} \times \RR{U_cY }}{\RR{AU_c} + \RR{U_cY } - 1} \times 1  \nonumber\\
& = x^2 \times \frac{x^2}{2x - 1} \times \frac{x^2}{2x - 1} \times \frac{x^2}{2x - 1} \nonumber \\
& = \frac{x^8}{(2x - 1)^3}
\end{align*}
and interpret with respect to the same parameters.

\hypertarget{implementation-in-r}{%
\section{Implementation in R}\label{implementation-in-r}}

We can use new functions from the R package \texttt{EValue}\textsuperscript{6} to either calculate the appropriate multiple bias E-value or to calculate a bound for the bias, given proposed parameters. The primary new functions in the package, \texttt{multi\_bound()} and \texttt{multi\_evalue()}, accept a set of biases (out of \texttt{confounding()}, \texttt{selection()}, and \texttt{misclassification()}, which take various arguments describing the bias in more detail). The function \texttt{multi\_bias()} is used to declare those biases. The \texttt{multi\_bound()} function requires values for the parameters making up the bound for the biases in question. The \texttt{multi\_evalue()} function requires just a value for the observed risk ratio, and prints a message to the user about the sensitivity parameters it refers to.

We will demonstrate the new package functionality by working through the examples in the main text. We will then show how the new functions can be used to recreate examples from earlier literature as well.

\begin{Shaded}
\begin{Highlighting}[]
\KeywordTok{library}\NormalTok{(EValue)}
\end{Highlighting}
\end{Shaded}

\hypertarget{examples-from-the-main-text}{%
\subsection{Examples from the main text}\label{examples-from-the-main-text}}

The \texttt{multi\_bias()} function takes as arguments one or more of the three bias functions, \texttt{confounding()}, \texttt{selection()}, and \texttt{misclassification()}. They should be listed in the order in which they occur in the data (i.e., does the measurement happen in the sample, or is the sample selected based on mismeasured exposure or outcome values?). Each of \texttt{selection()} and \texttt{misclassification()} take additional arguments depending on the assumptions and simplifications of a given scenario.

In the HIV example, we were interested in the composite bias due to confounding and selection. We were willing to make the assumption that the outcome is more likely in the selected portion of both exposure groups, so we include the argument \texttt{\ditto increased\ risk\ditto }. (The \texttt{\ditto general\ditto } argument is in contrast to \texttt{\ditto selected\ditto }, the latter meaning that we are only interested in inference in the selected population. Since \texttt{\ditto general\ditto } is the default, we could leave it out.)

\begin{Shaded}
\begin{Highlighting}[]
\NormalTok{HIV_biases <-}\StringTok{ }\KeywordTok{multi_bias}\NormalTok{(}\KeywordTok{confounding}\NormalTok{(), }
                         \KeywordTok{selection}\NormalTok{(}\StringTok{"general"}\NormalTok{, }\StringTok{"increased risk"}\NormalTok{))}
\end{Highlighting}
\end{Shaded}

Printing the biases prints out the arguments that are required for the \texttt{multi\_bound()} function for easy copying and pasting into that function.

\begin{Shaded}
\begin{Highlighting}[]
\NormalTok{HIV_biases}
\end{Highlighting}
\end{Shaded}

\begin{verbatim}
The following arguments can be copied and pasted into the multi_bound() function:
\end{verbatim}

\begin{verbatim}
RRAUc = , RRUcY = , RRUsYA1 = , RRSUsA1  =
\end{verbatim}

\begin{Shaded}
\begin{Highlighting}[]
\KeywordTok{multi_bound}\NormalTok{(}\DataTypeTok{biases =}\NormalTok{ HIV_biases, }
            \DataTypeTok{RRAUc =} \FloatTok{2.3}\NormalTok{, }\DataTypeTok{RRUcY =} \FloatTok{2.5}\NormalTok{, }\DataTypeTok{RRUsYA1 =} \DecValTok{3}\NormalTok{, }\DataTypeTok{RRSUsA1 =} \DecValTok{2}\NormalTok{)}
\end{Highlighting}
\end{Shaded}

\begin{verbatim}
[1] 2.269737
\end{verbatim}

Because the labeling of the arguments is not necessarily intuitive, we might want to confirm which refers to which parameter. We can use the \texttt{summary()} function on a object created with the \texttt{multi\_bias()} function to print more information about the biases.

\begin{Shaded}
\begin{Highlighting}[]
\KeywordTok{summary}\NormalTok{(HIV_biases)}
\end{Highlighting}
\end{Shaded}

\begin{verbatim}
         bias     output argument
1 confounding     RR_AUc    RRAUc
2 confounding     RR_UcY    RRUcY
3   selection RR_UsY|A=1  RRUsYA1
4   selection RR_SUs|A=1  RRSUsA1
\end{verbatim}

For easy copying and pasting of the notation we used in this appendix and in the main text, the argument \texttt{latex\ =\ TRUE} can be used in the summary function to print out an additional column with the parameters in our notation.

To calculate a multi-bias E-value, we must provide the observed effect estimate along with the set of biases. There are two options for doing so. The first is to declare the effect estimate with one of \texttt{RR()}, \texttt{OR()}, or \texttt{HR()}, depending on whether it is a risk, odds, or hazard ratio.

\begin{Shaded}
\begin{Highlighting}[]
\KeywordTok{multi_evalue}\NormalTok{(}\DataTypeTok{biases =}\NormalTok{ HIV_biases, }
             \DataTypeTok{est =} \KeywordTok{OR}\NormalTok{(}\FloatTok{6.75}\NormalTok{, }\DataTypeTok{rare =} \OtherTok{TRUE}\NormalTok{), }
             \DataTypeTok{lo =} \FloatTok{2.79}\NormalTok{, }\DataTypeTok{hi =} \FloatTok{16.31}\NormalTok{)}
\end{Highlighting}
\end{Shaded}

\begin{verbatim}
This multi-bias e-value refers simultaneously to parameters RRAUc, RRUcY, RRYAa. (See documentation for details.)
\end{verbatim}

\begin{verbatim}
                       point    lower upper
RR                  6.750000 2.790000 16.31
Multi-bias e-values 4.635703 2.728474    NA
\end{verbatim}

The lower and upper bound of the confidence interval are assumed to be on the same scale.

Alternatively, we can specify the scale of effect estimate using the \texttt{measure\ =} argument.

\begin{Shaded}
\begin{Highlighting}[]
\KeywordTok{multi_evalue}\NormalTok{(}\DataTypeTok{biases =}\NormalTok{ HIV_biases, }
             \DataTypeTok{est =} \FloatTok{6.75}\NormalTok{, }\DataTypeTok{measure =} \StringTok{"OR"}\NormalTok{, }\DataTypeTok{rare =} \OtherTok{TRUE}\NormalTok{, }
             \DataTypeTok{lo =} \FloatTok{2.79}\NormalTok{, }\DataTypeTok{hi =} \FloatTok{16.31}\NormalTok{)}
\end{Highlighting}
\end{Shaded}

\begin{verbatim}
This multi-bias e-value refers simultaneously to parameters RRAUc, RRUcY, RRYAa. (See documentation for details.)
\end{verbatim}

\begin{verbatim}
                       point    lower upper
RR                  6.750000 2.790000 16.31
Multi-bias e-values 4.635703 2.728474    NA
\end{verbatim}

Next we will look at the vitamins-leukemia example from the text. The \texttt{misclassification()} bias requires one of either \texttt{\ditto outcome\ditto } or \texttt{\ditto exposure\ditto }; if exposure misclassification is of interest, the user is also required to specify whether the outcome and/or exposure are sufficiently rare to use a risk ratio approximation for an odds ratio (irrespective of whether the effect estimate is actually on the odds ratio scale).

\begin{Shaded}
\begin{Highlighting}[]
\NormalTok{leuk_biases <-}\StringTok{ }\KeywordTok{multi_bias}\NormalTok{(}\KeywordTok{confounding}\NormalTok{(), }
                          \KeywordTok{misclassification}\NormalTok{(}\StringTok{"exposure"}\NormalTok{, }
                                            \DataTypeTok{rare_outcome =} \OtherTok{TRUE}\NormalTok{, }
                                            \DataTypeTok{rare_exposure =} \OtherTok{FALSE}\NormalTok{))}
\NormalTok{leuk_biases}
\end{Highlighting}
\end{Shaded}

\begin{verbatim}
The following arguments can be copied and pasted into the multi_bound() function:
\end{verbatim}

\begin{verbatim}
RRAUc = , RRUcY = , ORYAa  =
\end{verbatim}

Again we can calculate the bound and multi-bias E-value as in the text.

\begin{Shaded}
\begin{Highlighting}[]
\KeywordTok{multi_bound}\NormalTok{(}\DataTypeTok{biases =}\NormalTok{ leuk_biases, }\DataTypeTok{RRAUc =} \DecValTok{2}\NormalTok{, }\DataTypeTok{RRUcY =} \FloatTok{1.22}\NormalTok{, }\DataTypeTok{ORYAa =} \FloatTok{1.59}\NormalTok{) }
\end{Highlighting}
\end{Shaded}

\begin{verbatim}
[1] 1.747568
\end{verbatim}

\begin{Shaded}
\begin{Highlighting}[]
\KeywordTok{multi_evalue}\NormalTok{(}\DataTypeTok{biases =}\NormalTok{ leuk_biases, }
             \DataTypeTok{est =} \KeywordTok{OR}\NormalTok{(}\FloatTok{0.51}\NormalTok{, }\DataTypeTok{rare =} \OtherTok{TRUE}\NormalTok{), }
             \DataTypeTok{lo =} \FloatTok{0.3}\NormalTok{, }\DataTypeTok{hi =} \FloatTok{0.89}\NormalTok{)}
\end{Highlighting}
\end{Shaded}

\begin{verbatim}
This multi-bias e-value refers simultaneously to parameters RRAUc, RRUcY, RRYAa. (See documentation for details.)
\end{verbatim}

\begin{verbatim}
                       point lower    upper
RR                  0.510000   0.3 0.890000
Multi-bias e-values 1.351985    NA 1.058404
\end{verbatim}

We can easily demonstrate that the E-value is the same whether or not the effect estimate is inverted if the exposure is apparently protective. Also, if we don't want the message about the parameters to print, we can use the argument \texttt{message\ =\ FALSE}.

\begin{Shaded}
\begin{Highlighting}[]
\KeywordTok{multi_evalue}\NormalTok{(}\DataTypeTok{biases =}\NormalTok{ leuk_biases, }
             \DataTypeTok{est =} \KeywordTok{OR}\NormalTok{(}\DecValTok{1}\OperatorTok{/}\FloatTok{0.51}\NormalTok{, }\DataTypeTok{rare =} \OtherTok{TRUE}\NormalTok{), }
             \DataTypeTok{hi =} \DecValTok{1}\OperatorTok{/}\FloatTok{0.3}\NormalTok{, }\DataTypeTok{lo =} \DecValTok{1}\OperatorTok{/}\FloatTok{0.89}\NormalTok{,}
             \DataTypeTok{message =} \OtherTok{FALSE}\NormalTok{)}
\end{Highlighting}
\end{Shaded}

\begin{verbatim}
This multi-bias e-value refers simultaneously to parameters RRAUc, RRUcY, RRYAa. (See documentation for details.)
\end{verbatim}

\begin{verbatim}
                       point    lower    upper
RR                  1.960784 1.123596 3.333333
Multi-bias e-values 1.351985 1.058404       NA
\end{verbatim}

Finally, we presented a multi-bias E-value for all three biases. We can use the \texttt{summary()} function to just print the single value, instead of the matrix of the estimates and confidence limits and E-values for both.

\begin{Shaded}
\begin{Highlighting}[]
\KeywordTok{summary}\NormalTok{(}\KeywordTok{multi_evalue}\NormalTok{(}\DataTypeTok{biases =} \KeywordTok{multi_bias}\NormalTok{(}\KeywordTok{confounding}\NormalTok{(),}
                                         \KeywordTok{selection}\NormalTok{(}\StringTok{"general"}\NormalTok{),}
                                         \KeywordTok{misclassification}\NormalTok{(}\StringTok{"outcome"}\NormalTok{)),}
                     \DataTypeTok{est =} \KeywordTok{RR}\NormalTok{(}\DecValTok{4}\NormalTok{)))}
\end{Highlighting}
\end{Shaded}

\begin{verbatim}
[1] 1.888478
\end{verbatim}

\hypertarget{extensions-not-appearing-in-the-main-text}{%
\subsubsection{Extensions not appearing in the main text}\label{extensions-not-appearing-in-the-main-text}}

We may want to vary the magnitude of the parameters used to calculate the bounds. We'll use the biases from the HIV example to demonstrate.

\begin{Shaded}
\begin{Highlighting}[]
\CommentTok{# original bound}
\KeywordTok{multi_bound}\NormalTok{(}\DataTypeTok{biases =}\NormalTok{ HIV_biases, }\DataTypeTok{RRAUc =} \DecValTok{2}\NormalTok{, }\DataTypeTok{RRUcY =} \FloatTok{2.5}\NormalTok{, }
            \DataTypeTok{RRUsYA1 =} \DecValTok{3}\NormalTok{, }\DataTypeTok{RRSUsA1 =} \DecValTok{2}\NormalTok{)}
\end{Highlighting}
\end{Shaded}

\begin{verbatim}
[1] 2.142857
\end{verbatim}

\begin{Shaded}
\begin{Highlighting}[]
\CommentTok{# vary RRAUc from 1.25 to 3}
\KeywordTok{sapply}\NormalTok{(}\KeywordTok{seq}\NormalTok{(}\FloatTok{1.25}\NormalTok{, }\DecValTok{3}\NormalTok{, }\DataTypeTok{by =} \FloatTok{.25}\NormalTok{), }\ControlFlowTok{function}\NormalTok{(RRAUc) \{}
  \KeywordTok{multi_bound}\NormalTok{(}\DataTypeTok{biases =}\NormalTok{ HIV_biases, }\DataTypeTok{RRAUc =}\NormalTok{ RRAUc, }
              \DataTypeTok{RRUcY =} \FloatTok{2.5}\NormalTok{, }\DataTypeTok{RRUsYA1 =} \DecValTok{3}\NormalTok{, }\DataTypeTok{RRSUsA1 =} \DecValTok{2}\NormalTok{)}
\NormalTok{  \})}
\end{Highlighting}
\end{Shaded}

\begin{verbatim}
[1] 1.704545 1.875000 2.019231 2.142857 2.250000 2.343750 2.426471 2.500000
\end{verbatim}

\begin{Shaded}
\begin{Highlighting}[]
\CommentTok{# vary RRAUc and RRUcY}
\NormalTok{param_vals <-}\StringTok{ }\KeywordTok{seq}\NormalTok{(}\FloatTok{1.25}\NormalTok{, }\DecValTok{3}\NormalTok{, }\DataTypeTok{by =} \FloatTok{.25}\NormalTok{)}

\NormalTok{params <-}\StringTok{ }\KeywordTok{expand.grid}\NormalTok{(}\DataTypeTok{RRAUc =}\NormalTok{ param_vals,}
                      \DataTypeTok{RRUcY =}\NormalTok{ param_vals)}

\NormalTok{vals <-}\StringTok{ }\KeywordTok{mapply}\NormalTok{(multi_bound, }
               \DataTypeTok{RRAUc =}\NormalTok{ params}\OperatorTok{$}\NormalTok{RRAUc, }
               \DataTypeTok{RRUcY =}\NormalTok{ params}\OperatorTok{$}\NormalTok{RRUcY, }
               \DataTypeTok{MoreArgs =} \KeywordTok{list}\NormalTok{(}\DataTypeTok{biases =}\NormalTok{ HIV_biases, }
                               \DataTypeTok{RRUsYA1 =} \DecValTok{3}\NormalTok{, }\DataTypeTok{RRSUsA1 =} \DecValTok{2}\NormalTok{))}
\KeywordTok{matrix}\NormalTok{(vals,}
  \DataTypeTok{ncol =} \KeywordTok{length}\NormalTok{(param_vals),}
  \DataTypeTok{dimnames =} \KeywordTok{list}\NormalTok{(param_vals, param_vals)}
\NormalTok{)}
\end{Highlighting}
\end{Shaded}

\begin{verbatim}
         1.25      1.5     1.75        2     2.25      2.5     2.75        3
1.25 1.562500 1.607143 1.640625 1.666667 1.687500 1.704545 1.718750 1.730769
1.5  1.607143 1.687500 1.750000 1.800000 1.840909 1.875000 1.903846 1.928571
1.75 1.640625 1.750000 1.837500 1.909091 1.968750 2.019231 2.062500 2.100000
2    1.666667 1.800000 1.909091 2.000000 2.076923 2.142857 2.200000 2.250000
2.25 1.687500 1.840909 1.968750 2.076923 2.169643 2.250000 2.320312 2.382353
2.5  1.704545 1.875000 2.019231 2.142857 2.250000 2.343750 2.426471 2.500000
2.75 1.718750 1.903846 2.062500 2.200000 2.320312 2.426471 2.520833 2.605263
3    1.730769 1.928571 2.100000 2.250000 2.382353 2.500000 2.605263 2.700000
\end{verbatim}

Of course, all of the parameters in the bound could be varied, but summarizing the resulting bounds in a simple table or figure becomes more difficult with more than two dimensions.

When calculating a multi-bias E-value, we may also think that the null is unlikely but wish to consider how much bias could have shifted a different true value to the observed value. For example, in the HIV example, we could calculate a multi-bias E-value for a true risk ratio of 2 rather than the null value of 1:

\begin{Shaded}
\begin{Highlighting}[]
\KeywordTok{multi_evalue}\NormalTok{(}\DataTypeTok{biases =}\NormalTok{ HIV_biases, }
             \DataTypeTok{est =} \KeywordTok{OR}\NormalTok{(}\FloatTok{6.75}\NormalTok{, }\DataTypeTok{rare =} \OtherTok{TRUE}\NormalTok{), }
             \DataTypeTok{lo =} \FloatTok{2.79}\NormalTok{, }\DataTypeTok{hi =} \FloatTok{16.31}\NormalTok{,}
             \DataTypeTok{true =} \DecValTok{2}\NormalTok{)}
\end{Highlighting}
\end{Shaded}

\begin{verbatim}
You are calculating a "non-null" multi-bias E-value, i.e., a multi-bias E-value for the minimum amount of bias needed to move the estimate and confidence interval to your specified true value of 2 rather than to the null value.
\end{verbatim}

\begin{verbatim}
This multi-bias e-value refers simultaneously to parameters RRAUc, RRUcY, RRYAa. (See documentation for details.)
\end{verbatim}

\begin{verbatim}
                       point    lower upper
RR                  6.750000 2.790000 16.31
Multi-bias e-values 3.077243 1.643623    NA
\end{verbatim}

The multi-bias E-value for the point estimate, 3.08 is of course smaller than the \enquote{null} E-value of 4.64, as less bias could have resulted in an OR of 6.75 if the true OR were 2 than would have been necessary to shift it from 1.

The interpretation of the parameters differs depending on the ordering of the selection bias and misclassification. We can see that the parameters expected in the \texttt{multi\_bound()} function and printed by the \texttt{multi\_evalue()} function reflect the ordering in which the biases are added to \texttt{multi\_bias()} (see \texttt{output} column).

\begin{Shaded}
\begin{Highlighting}[]
\CommentTok{# misclassification occurs in the selected group}
\KeywordTok{summary}\NormalTok{(}
  \KeywordTok{multi_bias}\NormalTok{(}\KeywordTok{selection}\NormalTok{(}\StringTok{"general"}\NormalTok{),}
             \KeywordTok{misclassification}\NormalTok{(}\StringTok{"exposure"}\NormalTok{, }\DataTypeTok{rare_outcome =} \OtherTok{TRUE}\NormalTok{))}
\NormalTok{  )}
\end{Highlighting}
\end{Shaded}

\begin{verbatim}
                        bias     output argument
1                  selection RR_UsY|A=1  RRUsYA1
2                  selection RR_SUs|A=1  RRSUsA1
3                  selection RR_UsY|A=0  RRUsYA0
4                  selection RR_SUs|A=0  RRSUsA0
5 exposure misclassification OR_YA*|a,S   ORYAaS
\end{verbatim}

\begin{Shaded}
\begin{Highlighting}[]
\CommentTok{# selection is of misclassified individuals}
\KeywordTok{summary}\NormalTok{(}
  \KeywordTok{multi_bias}\NormalTok{(}\KeywordTok{misclassification}\NormalTok{(}\StringTok{"exposure"}\NormalTok{, }\DataTypeTok{rare_outcome =} \OtherTok{TRUE}\NormalTok{),}
             \KeywordTok{selection}\NormalTok{(}\StringTok{"general"}\NormalTok{))}
\NormalTok{  )}
\end{Highlighting}
\end{Shaded}

\begin{verbatim}
                        bias      output argument
1                  selection RR_UsY|A*=0  RRUsYA0
2                  selection RR_SUs|A*=1  RRSUsA1
3                  selection RR_UsY|A*=1  RRUsYA1
4                  selection RR_SUs|A*=1  RRSUsA1
5 exposure misclassification    OR_YA*|a    ORYAa
\end{verbatim}

When selection bias and confounding are both of interest, but restricting inference to the selected population only is desired, the parameters are shared by the two biases:

\begin{Shaded}
\begin{Highlighting}[]
\KeywordTok{summary}\NormalTok{(}
  \KeywordTok{multi_bias}\NormalTok{(}\KeywordTok{confounding}\NormalTok{(),}
             \KeywordTok{selection}\NormalTok{(}\StringTok{"selected"}\NormalTok{),}
             \KeywordTok{misclassification}\NormalTok{(}\StringTok{"exposure"}\NormalTok{, }\DataTypeTok{rare_outcome =} \OtherTok{TRUE}\NormalTok{))}
\NormalTok{  )}
\end{Highlighting}
\end{Shaded}

\begin{verbatim}
                        bias     output argument
1  confounding and selection  RR_AUsc|S  RRAUscS
2  confounding and selection  RR_UscY|S  RRUscYS
3 exposure misclassification OR_YA*|a,S   ORYAaS
\end{verbatim}

Finally, we can see the expected relationship between the multi-bias bound and the multi-bias E-value.

\begin{Shaded}
\begin{Highlighting}[]
\NormalTok{biases <-}\StringTok{ }\KeywordTok{multi_bias}\NormalTok{(}\KeywordTok{confounding}\NormalTok{(), }
                     \KeywordTok{selection}\NormalTok{(}\StringTok{"general"}\NormalTok{, }\StringTok{"decreased risk"}\NormalTok{), }
                     \KeywordTok{misclassification}\NormalTok{(}\StringTok{"outcome"}\NormalTok{))}

\CommentTok{# calculate bound with those parameters all equal to 2}
\KeywordTok{multi_bound}\NormalTok{(biases, }\DataTypeTok{RRAUc =} \DecValTok{2}\NormalTok{, }\DataTypeTok{RRUcY =} \DecValTok{2}\NormalTok{, }\DataTypeTok{RRUsYA0 =} \DecValTok{2}\NormalTok{, }\DataTypeTok{RRSUsA0 =} \DecValTok{2}\NormalTok{, }\DataTypeTok{RRAYyS =} \DecValTok{2}\NormalTok{)}
\end{Highlighting}
\end{Shaded}

\begin{verbatim}
[1] 3.555556
\end{verbatim}

\begin{Shaded}
\begin{Highlighting}[]
\CommentTok{# get multi-bias e-value for that value; should be ~2}
\KeywordTok{summary}\NormalTok{(}\KeywordTok{multi_evalue}\NormalTok{(biases, }\DataTypeTok{est =} \KeywordTok{RR}\NormalTok{(}\FloatTok{3.555556}\NormalTok{)))}
\end{Highlighting}
\end{Shaded}

\begin{verbatim}
[1] 1.999997
\end{verbatim}

\hypertarget{examples-from-earlier-literature}{%
\subsection{Examples from earlier literature}\label{examples-from-earlier-literature}}

The multi-bias bound and E-value are generalizations of previously published results. To demonstrate, we recreate here some examples from three articles introducing the bound and E-value concept for confounding, selection bias, and differential misclassification.

\hypertarget{from-sensitivity-analysis-without-assumptions-ding-vanderweele-2016ding2016b}{%
\subsubsection{\texorpdfstring{From \emph{Sensitivity Analysis without Assumptions}, Ding \& VanderWeele 2016\textsuperscript{3}}{From Sensitivity Analysis without Assumptions, Ding \& VanderWeele 20163}}\label{from-sensitivity-analysis-without-assumptions-ding-vanderweele-2016ding2016b}}

\begin{Shaded}
\begin{Highlighting}[]
\CommentTok{# example from page 370}
\NormalTok{biases_ex1 <-}\StringTok{ }\KeywordTok{confounding}\NormalTok{()}
\CommentTok{# specifying parameters in bound}
\KeywordTok{multi_bound}\NormalTok{(}\DataTypeTok{biases =}\NormalTok{ biases_ex1, }\DataTypeTok{RRAUc =} \DecValTok{2}\NormalTok{, }\DataTypeTok{RRUcY =} \DecValTok{2}\NormalTok{)}
\end{Highlighting}
\end{Shaded}

\begin{verbatim}
[1] 1.333333
\end{verbatim}

\begin{Shaded}
\begin{Highlighting}[]
\CommentTok{# Table 1, page 371}
\CommentTok{# consider all possible combinations for bound}
\NormalTok{param_vals <-}\StringTok{ }\KeywordTok{c}\NormalTok{(}\FloatTok{1.3}\NormalTok{, }\FloatTok{1.5}\NormalTok{, }\FloatTok{1.8}\NormalTok{, }\DecValTok{2}\NormalTok{, }\FloatTok{2.5}\NormalTok{, }\DecValTok{3}\NormalTok{, }\FloatTok{3.5}\NormalTok{, }\DecValTok{4}\NormalTok{, }\DecValTok{5}\NormalTok{, }\DecValTok{6}\NormalTok{, }\DecValTok{8}\NormalTok{, }\DecValTok{10}\NormalTok{)}
\NormalTok{params <-}\StringTok{ }\KeywordTok{expand.grid}\NormalTok{(}\DataTypeTok{RRAUc =}\NormalTok{ param_vals,}
                      \DataTypeTok{RRUcY =}\NormalTok{ param_vals)}
\NormalTok{table1_vals <-}\StringTok{ }\KeywordTok{mapply}\NormalTok{(multi_bound, }\DataTypeTok{RRAUc =}\NormalTok{ params}\OperatorTok{$}\NormalTok{RRAUc, }\DataTypeTok{RRUcY =}\NormalTok{ params}\OperatorTok{$}\NormalTok{RRUcY, }
                      \DataTypeTok{MoreArgs =} \KeywordTok{list}\NormalTok{(}\DataTypeTok{biases =}\NormalTok{ biases_ex1))}
\NormalTok{table1 <-}\StringTok{ }\KeywordTok{matrix}\NormalTok{(table1_vals,}
  \DataTypeTok{ncol =} \KeywordTok{length}\NormalTok{(param_vals),}
  \DataTypeTok{dimnames =} \KeywordTok{list}\NormalTok{(param_vals, param_vals)}
\NormalTok{)}
\KeywordTok{round}\NormalTok{(table1, }\DecValTok{2}\NormalTok{)}
\end{Highlighting}
\end{Shaded}

\begin{verbatim}
     1.3  1.5  1.8    2  2.5    3  3.5    4    5    6    8   10
1.3 1.06 1.08 1.11 1.13 1.16 1.18 1.20 1.21 1.23 1.24 1.25 1.26
1.5 1.08 1.12 1.17 1.20 1.25 1.29 1.31 1.33 1.36 1.38 1.41 1.43
1.8 1.11 1.17 1.25 1.29 1.36 1.42 1.47 1.50 1.55 1.59 1.64 1.67
2   1.13 1.20 1.29 1.33 1.43 1.50 1.56 1.60 1.67 1.71 1.78 1.82
2.5 1.16 1.25 1.36 1.43 1.56 1.67 1.75 1.82 1.92 2.00 2.11 2.17
3   1.18 1.29 1.42 1.50 1.67 1.80 1.91 2.00 2.14 2.25 2.40 2.50
3.5 1.20 1.31 1.47 1.56 1.75 1.91 2.04 2.15 2.33 2.47 2.67 2.80
4   1.21 1.33 1.50 1.60 1.82 2.00 2.15 2.29 2.50 2.67 2.91 3.08
5   1.23 1.36 1.55 1.67 1.92 2.14 2.33 2.50 2.78 3.00 3.33 3.57
6   1.24 1.38 1.59 1.71 2.00 2.25 2.47 2.67 3.00 3.27 3.69 4.00
8   1.25 1.41 1.64 1.78 2.11 2.40 2.67 2.91 3.33 3.69 4.27 4.71
10  1.26 1.43 1.67 1.82 2.17 2.50 2.80 3.08 3.57 4.00 4.71 5.26
\end{verbatim}

\begin{Shaded}
\begin{Highlighting}[]
\CommentTok{# reduce an observed RR of 2.5 to true value of 1.5, page 371}
\KeywordTok{summary}\NormalTok{(}\KeywordTok{multi_evalue}\NormalTok{(}\DataTypeTok{biases =} \KeywordTok{confounding}\NormalTok{(), }\DataTypeTok{est =} \KeywordTok{RR}\NormalTok{(}\FloatTok{2.5}\NormalTok{), }\DataTypeTok{true =} \FloatTok{1.5}\NormalTok{))}
\end{Highlighting}
\end{Shaded}

\begin{verbatim}
You are calculating a "non-null" multi-bias E-value, i.e., a multi-bias E-value for the minimum amount of bias needed to move the estimate and confidence interval to your specified true value of 1.5 rather than to the null value.
\end{verbatim}

\begin{verbatim}
[1] 2.720763
\end{verbatim}

\begin{Shaded}
\begin{Highlighting}[]
\CommentTok{# smoking and lung cancer e-value, page 373}
\KeywordTok{summary}\NormalTok{(}\KeywordTok{multi_evalue}\NormalTok{(}\DataTypeTok{biases =} \KeywordTok{confounding}\NormalTok{(), }\DataTypeTok{est =} \KeywordTok{RR}\NormalTok{(}\FloatTok{10.73}\NormalTok{)))}
\end{Highlighting}
\end{Shaded}

\begin{verbatim}
[1] 20.94777
\end{verbatim}

\hypertarget{from-bounding-bias-due-to-selection-smith-vanderweele-2019smith2019}{%
\subsubsection{\texorpdfstring{From \emph{Bounding bias due to selection}, Smith \& VanderWeele, 2019\textsuperscript{2}}{From Bounding bias due to selection, Smith \& VanderWeele, 20192}}\label{from-bounding-bias-due-to-selection-smith-vanderweele-2019smith2019}}

\begin{Shaded}
\begin{Highlighting}[]
\NormalTok{biases_ex2 <-}\StringTok{ }\KeywordTok{selection}\NormalTok{(}\StringTok{"general"}\NormalTok{)}

\CommentTok{# result 1A example}
\KeywordTok{multi_bound}\NormalTok{(}\DataTypeTok{biases =}\NormalTok{ biases_ex2,}
            \DataTypeTok{RRUsYA1 =} \DecValTok{2}\NormalTok{, }\DataTypeTok{RRSUsA1 =} \FloatTok{1.7}\NormalTok{, }\DataTypeTok{RRUsYA0 =} \DecValTok{2}\NormalTok{, }\DataTypeTok{RRSUsA0 =} \FloatTok{1.5}\NormalTok{)}
\end{Highlighting}
\end{Shaded}

\begin{verbatim}
[1] 1.511111
\end{verbatim}

\begin{Shaded}
\begin{Highlighting}[]
\CommentTok{# result 1B example}
\KeywordTok{multi_evalue}\NormalTok{(}\DataTypeTok{biases =}\NormalTok{ biases_ex2, }\DataTypeTok{est =} \KeywordTok{OR}\NormalTok{(}\FloatTok{73.1}\NormalTok{, }\DataTypeTok{rare =} \OtherTok{TRUE}\NormalTok{), }\DataTypeTok{lo =} \FloatTok{13.0}\NormalTok{)}
\end{Highlighting}
\end{Shaded}

\begin{verbatim}
This multi-bias e-value refers simultaneously to parameters RRAUc, RRUcY, RRYAa. (See documentation for details.)
\end{verbatim}

\begin{verbatim}
                       point     lower upper
RR                  73.10000 13.000000    NA
Multi-bias e-values 16.58415  6.670587    NA
\end{verbatim}

\begin{Shaded}
\begin{Highlighting}[]
\CommentTok{# result 4B example}
\KeywordTok{summary}\NormalTok{(}\KeywordTok{multi_evalue}\NormalTok{(}\DataTypeTok{biases =} \KeywordTok{selection}\NormalTok{(}\StringTok{"general"}\NormalTok{, }\StringTok{"S = U"}\NormalTok{, }\StringTok{"increased risk"}\NormalTok{), }
             \DataTypeTok{est =} \KeywordTok{OR}\NormalTok{(}\FloatTok{5.2}\NormalTok{, }\DataTypeTok{rare =} \OtherTok{TRUE}\NormalTok{)))}
\end{Highlighting}
\end{Shaded}

\begin{verbatim}
[1] 5.2
\end{verbatim}

\begin{Shaded}
\begin{Highlighting}[]
\CommentTok{# result 5B example}
\KeywordTok{multi_evalue}\NormalTok{(}\DataTypeTok{biases =} \KeywordTok{selection}\NormalTok{(}\StringTok{"selected"}\NormalTok{), }
             \DataTypeTok{est =} \KeywordTok{OR}\NormalTok{(}\FloatTok{1.5}\NormalTok{, }\DataTypeTok{rare =} \OtherTok{TRUE}\NormalTok{), }\DataTypeTok{lo =} \FloatTok{1.22}\NormalTok{)}
\end{Highlighting}
\end{Shaded}

\begin{verbatim}
This multi-bias e-value refers simultaneously to parameters RRAUc, RRUcY, RRYAa. (See documentation for details.)
\end{verbatim}

\begin{verbatim}
                       point    lower upper
RR                  1.500000 1.220000    NA
Multi-bias e-values 2.366025 1.738081    NA
\end{verbatim}

\hypertarget{from-simple-sensitivity-analysis-for-differential-measurement-error-vanderweele-li-2019vanderweele2019b}{%
\subsubsection{\texorpdfstring{From \emph{Simple Sensitivity Analysis for Differential Measurement Error}, VanderWeele \& Li 2019\textsuperscript{1}}{From Simple Sensitivity Analysis for Differential Measurement Error, VanderWeele \& Li 20191}}\label{from-simple-sensitivity-analysis-for-differential-measurement-error-vanderweele-li-2019vanderweele2019b}}

\begin{Shaded}
\begin{Highlighting}[]
\NormalTok{biases_ex3 <-}\StringTok{ }\KeywordTok{misclassification}\NormalTok{(}\StringTok{"exposure"}\NormalTok{, }
                                \DataTypeTok{rare_outcome =} \OtherTok{TRUE}\NormalTok{, }\DataTypeTok{rare_exposure =} \OtherTok{TRUE}\NormalTok{)}
\KeywordTok{multi_evalue}\NormalTok{(}\DataTypeTok{biases =}\NormalTok{ biases_ex3, }\DataTypeTok{est =} \KeywordTok{OR}\NormalTok{(}\FloatTok{1.51}\NormalTok{, }\DataTypeTok{rare =} \OtherTok{TRUE}\NormalTok{), }\DataTypeTok{lo =} \FloatTok{1.03}\NormalTok{)}
\end{Highlighting}
\end{Shaded}

\begin{verbatim}
This multi-bias e-value refers simultaneously to parameters RRAUc, RRUcY, RRYAa. (See documentation for details.)
\end{verbatim}

\begin{verbatim}
                    point lower upper
RR                   1.51  1.03    NA
Multi-bias e-values  1.51  1.03    NA
\end{verbatim}

\newpage

\hypertarget{more-examples-of-dags-for-multiple-biases}{%
\section{More examples of DAGs for multiple biases}\label{more-examples-of-dags-for-multiple-biases}}

These examples show how various combinations of biases can be represented by directed acyclic graphs, and the independence assumptions that are implied.

\begin{figure}[h]

{\centering \includegraphics{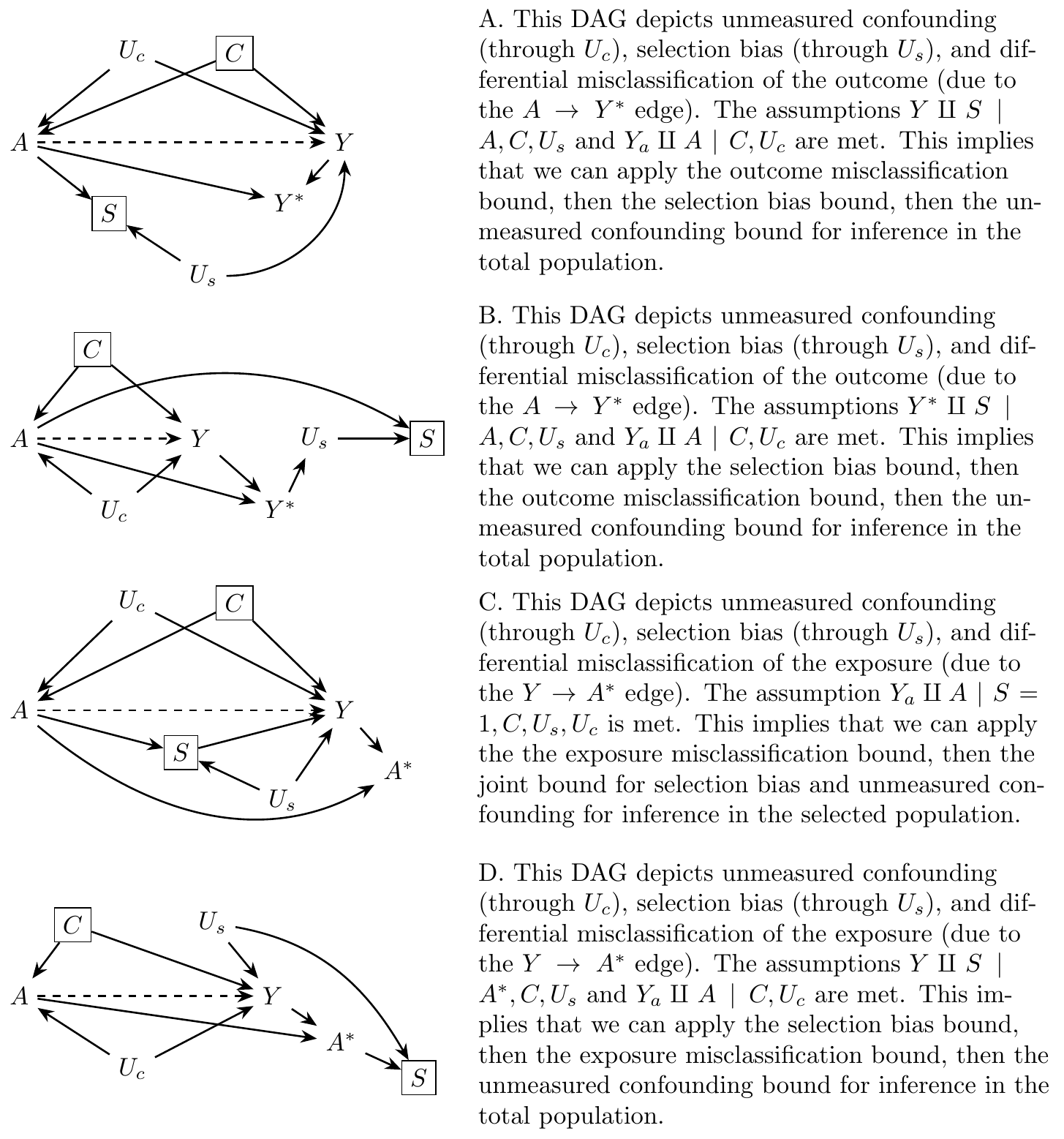} 

}

\caption{Directed acyclic graphs depicting multiple biases.}\label{fig:DAGs}
\end{figure}

\newpage

\hypertarget{references2}{%
\section*{References}\label{references2}}
\addcontentsline{toc}{section}{References}

\hypertarget{refs}{}
\leavevmode\hypertarget{ref-VanderWeele2019b}{}%
1. VanderWeele TJ, Li Y. Simple sensitivity analysis for differential measurement error. \emph{Am J Epidemiol}. 2019;188:1823--1829.

\leavevmode\hypertarget{ref-Smith2019}{}%
2. Smith LH, VanderWeele TJ. Bounding bias due to selection. \emph{Epidemiology}. 2019;30:509--516.

\leavevmode\hypertarget{ref-Ding2016b}{}%
3. Ding P, VanderWeele TJ. Sensitivity analysis without assumptions. \emph{Epidemiology}. 2016;27:368--377.

\leavevmode\hypertarget{ref-VanderWeele2019d}{}%
4. VanderWeele TJ. Optimal approximate conversions of odds ratios and hazard ratios to risk ratios. \emph{Biometrics}. 2019;1--7.

\leavevmode\hypertarget{ref-VanderWeele2017a}{}%
5. VanderWeele TJ, Ding P. Sensitivity analysis in observational research: Introducing the e-value. \emph{Ann Intern Med}. 2017;167:268--275.

\leavevmode\hypertarget{ref-Mathur2018}{}%
6. Mathur MB, Ding P, Riddell CA, et al. Website and r package for computing e-values. \emph{Epidemiology}. 2018;29:e45--e47.

\end{document}